\documentclass[useAMS]{mn2e}
\usepackage{psfig,epsfig}
\usepackage{graphicx}

\begin{document}
\def\simlt{\mathrel{\rlap{\lower 3pt\hbox{$\sim$}}
        \raise 2.0pt\hbox{$<$}}}
\def\simgt{\mathrel{\rlap{\lower 3pt\hbox{$\sim$}}
        \raise 2.0pt\hbox{$>$}}}

\title[The {\it Herschel}-PEP survey: evidence for downsizing in the hosts of dusty star-forming systems] {The {\it Herschel}-PEP survey: evidence for downsizing in the hosts of dusty star-forming systems}

\author[Manuela Magliocchetti et al.]
{\parbox[t]\textwidth{M. Magliocchetti$^{1}$, P. Popesso$^{2}$,  D. Rosario$^{2}$, D. Lutz$^{2}$, H. Aussel$^{3}$, S. Berta$^{2}$, B. Altieri$^{4}$, P. Andreani$^{5}$, J. Cepa$^{6}$, H. Casta\~neda$^{6}$, A. Cimatti$^{7}$,  D. Elbaz$^{3}$, R. Genzel$^{2}$,
    A. Grazian$^{8}$, C. Gruppioni$^{9}$, O. Ilbert$^{10}$, E. Le Floc'h$^{3}$,   B. Magnelli$^{2}$, R. Maiolino$^{8}$, 
    R. Nordon$^{2}$, A. Poglitsch$^{2}$,
      F. Pozzi$^{7}$, L. Riguccini$^{3}$, G. Rodighiero$^{11}$, M. Sanchez-Portal$^{4}$,  P. Santini$^{8}$
 N.M.F\"orster Schreiber$^{2}$, E. Sturm$^{2}$, L. Tacconi$^{2}$, I. Valtchanov$^{4}$}\\
 \\
{\tt $^1$ INAF-IAPS, Via Fosso del Cavaliere 100, 00133, Roma,
  Italy}\\
{\tt $^2$ Max Planck Institute f\"ur Extraterrestrische Physik (MPE),
  Postfach 1312,  D85741, Garching, Germany}\\
{\tt $^3$ CEA-Saclay, Service d'Astrophysique, F-91191,
  Gif-sur-Yvette, France}\\
{\tt $^4$ ESA, P.O. Box 78, 28691 Villanueva de la Ca\~nada, Madrid, Spain}\\
{\tt $^5$ ESO, Karl Schwarzschild Strasse 2, D85748, Garching, Germany}\\
{\tt $^6$ Instituto de Astrofisica de Canarias, V'a Lactea, E38205, La Laguna (Tenerife), Spain}\\
{\tt $^7$ Dipartimento di Astronomia, Universita' di Bologna, Via Ranzani 1, 40127, Bologna, Italy}\\
{\tt $^8$ INAF-Osservatorio Astronomico di Roma, Via di Frascati 33, 00040 Monte Porzio Catone, Italy}\\
{\tt $^{9}$ INAF-Osservatorio Astronomico di Bologna, Via Ranzani 1, 40127, Bologna, Italy}\\
{\tt $^{10}$  LAM-rue FrŽdŽric Joliot-Curie 38, 13388, Marseille cedex 13,  FRANCE}\\\
{\tt $^{11}$ Dipartimento di Astronomia, Universita' di Padova,
  Vicolo dell' Osservatorio 3, 35122, Padova, Italy}
} 
\maketitle

\begin{abstract}
By making use of {\it Herschel}-PEP observations of the COSMOS and Extended Groth Strip fields, we have estimated the dependence of the clustering properties of FIR-selected sources on their 100$\mu$m fluxes. 
Our analysis shows a tendency for the clustering strength to decrease with limiting fluxes. By assuming a power-law slope with $\gamma=1.8$ for the two-point correlation function $\xi(r)=(r/r_0)^{-\gamma}$, we find:        
$r_0(S_{100 \mu\rm{m}}\ge 8$ mJy$)=4.3^{+0.7 }_{-0.7}$ Mpc and $r_0(S_{100 \mu\rm{m}}\ge 5$ mJy$)=5.8^{+1.8 }_{-2.0}$. These values convert into minimum halo masses M$_{\rm min}\sim 10^{11.6}$ M$_\odot$ for sources brighter than 8 mJy and 
M$_{\rm min}\sim 10^{12.4}$ M$_\odot$ for fainter, $S_{100 \mu\rm{m}}\ge 5$ mJy galaxies. 
We show such an increase of the clustering strength to be due to an intervening population of $z\sim 2$ sources, which are very strongly clustered and whose relative contribution, equal to about 10\% of the total counts at $S_{100 \mu\rm{m}}\ge 2$ mJy, rapidly decreases for brighter flux cuts. By removing such a contribution, we find that 
$z\simlt 1$ FIR galaxies have approximately the same clustering properties, irrespective of their flux level.  The above results were then used to investigate the {\it intrinsic} dependence on cosmic epoch of the clustering strength of dusty star-forming galaxies between $z\sim 0$ and $z\sim 2.5$. This was done by comparing our dataset with IRAS in the local universe and with sources selected at 160$\mu$m in the GOODS-South at $z\simeq 2$.
In order to remove any bias in the selection process, the adopted sample only includes galaxies observed at  the same rest-frame wavelength, $\lambda\sim 60 \mu$m, which have comparable luminosities and therefore star-formation rates (SFR$\simgt 100$ M$_\odot$/yr). 
 Our analysis shows that the same amount of (intense) star forming activity takes place in extremely different environments at the different cosmological epochs. {\bf  For $z\simlt 1$} the hosts of such star forming systems are small, M$_{\rm min}\sim 10^{11}$ M$_\odot$, isolated galaxies.  High ($z\sim2$) redshift star formation instead seems to uniquely take place in extremely massive/cluster-like halos, M$_{\rm min}\sim 10^{13.5}$ M$_\odot$, which are associated with the highest peaks of the density fluctuation field at those epochs. 
 \end{abstract}

\section{Introduction}
Investigations of the Large Scale Structure (LSS) traced by selected classes of extra-galactic sources, combined with increasingly more refined models for its description, has enabled in the recent years the derivation of important information on some of the physical properties characterizing the objects which produce the clustering signal. One of these key quantities is the halo mass and its evolution with look back time, which for instance allows one to derive cosmological connections between different populations of extra-galactic sources observed in different wavelengths at different epochs. 

Within this framework, amongst the more interesting classes of sources are those which are undergoing some kind of active process such as AGN or star formation.
Since the first high redshift observations of the two Hubble Deep Fields (Williams et al. 1996; Casertano et al. 2000), all clustering analyses performed on optical/near-IR surveys which were at the same time wide enough to ensure statistical significance and deep enough to probe high redshift regimes, have highlighted a strong evolution of these two classes of sources, whereby activity shifts to smaller and smaller haloes/galaxies as one moves forward in time. 
The critical redshift for this transition seems to be around $z\sim 2$ for star-forming galaxies (e.g. Magliocchetti \& Maddox 1999; Arnouts et al. 2002; Foucaud et al. 2010; Hartley et al. 2010) and $z\simgt 3$ for QSOs (e.g. Porciani, Magliocchetti \& Norberg 2004; Shen et al. 2007; Ross et al. 2009; Shen et al. 2009). Perhaps not surprisingly, these two redshift values bracket the peak of cosmic activity both in terms of AGN and star formation (Madau et al. 1996).  \\
This pattern, referred to as "downsizing" (Cowie et al. 1996; Heavens et al. 2004), is also observed in an increasing number of independent works 
(see e.g. Treu et al. 2005; Cimatti, Daddi \& Renzini 2006, Saracco et al. 2006; Bundy et al. 2006; McLure et al. 2006 just to mention a few), all adopting techniques which differ from clustering analyses.

The advent of the {\it Spitzer} telescope has marked another milestone in our knowledge of the population of dusty (and therefore optically dim) active galaxies. 
In particular, for the first time Spitzer has detected a population of  Ultra-Luminous Infrared Galaxies (ULIRGS, $L\simgt 10^{12} L_\odot$) in the redshift range ($1.6\simlt z\simlt 2.5$) with extremely high clustering lengths ($r_0\sim 15$ Mpc -- Farrah et al. 2006; Magliocchetti et al. 2007; 2008; Brodwin et al. 2008; Starikova et al. 2012). Taken at face value, these correlation lengths in the local universe correspond to those of groups-to-poor clusters of galaxies (see e.g. Guzzo et al. 2000; Estrada et al. 2009), and might be interpreted with a roughly one-to-one correspondence between the overwhelming majority of ULIRGS at $z\sim 2$ and  $z\simlt 1$ clusters (e.g. Brodwin et al 2007).

The {\it Herschel} telescope allows the scientific community to further extend  its knowledge on the formation and evolution of dusty sources up to very high redshifts. First attempts to measure the clustering properties of  {\it Herschel}  galaxies have been presented by Maddox et al. (2010), Cooray et al. (2010) and Amblard et al. (2011) for  sources detected with the SPIRE (Griffin et al. 2010) instrument at the wavelengths $\lambda=250, 350$ and 500$\mu$m. However, the above estimates can at most be considered as tentative since in the first place they do not agree with each other, and most importantly, because the de-projected quantities which are reported in some of these works are solely based on models for the redshift distribution of SPIRE sources at the various wavelengths.

A step forward has been made possible by the {\it Herschel}-PEP survey, which provides deep observations taken with the PACS instrument (Poglitsch et al. 2010) at the wavelengths $\lambda=70, 100$ and 
160$\mu$m on a number of well studied fields such as COSMOS, GOODS-North and South, the Lockman region, the Extended Chandra Deep Field and the Extended Groth Strip. While probing less sources than those collected in the mainly SPIRE-based programs, there are undoubt advantages connected with the exploitation of PEP data: on the one hand, the shorter wavelength range implies a smaller angular resolution and therefore an easier search for counterparts of PEP-detected sources. On the other hand, observations of well known and studied regions such as the two GOODS fields or COSMOS, not only allow identification and characterization of  FIR galaxies at the different wavelengths, but also provide spectroscopic and/or phometric redshift determinations for the overwhelming majority of such sources.

By making use of PEP observations of the GOODS-South field, Magliocchetti et al. (2011) managed for the first time to provide direct (i.e. purely based on data) measurements of the three-dimensional clustering properties of far-infrared sources up to $z\sim 3$. These authors report a clustering length $r_0\simeq 6.5$ Mpc for a sample of galaxies with 100$\mu$m fluxes brighter than 2 mJy.  Possibly more importantly, they also find a tremendous increase of the clustering strength for galaxies residing at $z\sim 2$. Indeed, for this population the reported values for the correlation length are $r_0=17-19$ Mpc, which set these galaxies within very massive, cluster-scale dark matter halos. 

The results presented in Magliocchetti et al. (2011) are in agreement with those found in the literature for actively star-forming galaxies selected at redshifts $z\sim 2$ to $z\sim 3$ (cfr. beginning of this Introduction). 
However, although some steps forward have been taken in order to minimize its effect (see e.g. Magliocchetti et al. 2008; Focaud et al. 2010; Lin et al. 2012), all the conclusions on a possible {\it intrinsic} evolution of the clustering strength with look-back time are hampered by the well known 'Malmquist bias' issue, since in flux limited surveys the more distant sources are also the more luminous ones which, at least in the visible part of the spectrum,  are generally found to be intrinsically more clustered than fainter galaxies. Furthermore, a  survey which is carried out at a fixed wavelength/waveband will probe different portions of a source's electromagnetic spectrum at the different redshifts. As a result of this second effect,  a heterogeneous mixture of galaxy types might end up in a survey sample as one moves further back in time, preventing to draw any conclusion on the evolution of a specific population with redshift. 

The present paper aims at overcoming these two potential selection biases so to provide some information on the  {\it intrinsic} evolution of the clustering properties of FIR galaxies. This will be done by estimating the two-point correlation function $w(\theta)$ of PEP-selected galaxies with different fluxes. In order to do so, our analysis will include some of the deep fields probed by the PEP survey such as COSMOS, the Extended Groth Strip and GOODS-South.
The work is organized as follows: Section 2 introduces the PEP catalogues. Section 3 presents the results for the two-point correlation function of PEP sources as a function of their fluxes both in projected (\S 3.1) and in three-dimensional (\S3.2) quantities, while Section 4 provides some information on the physical properties of FIR galaxies such as the halo masses of their dark matter hosts. Section 5 uses the results found in \S 3 and \S 4 to investigate the clustering properties of relatively local,  {\bf $z\simlt 1$}, PEP-selected galaxies as a function of their flux. This sample will then be compared with others taken from the literature for sources selected at the same rest-frame wavelength as the ones considered in this work and with comparable bolometric luminosities, so to investigate the evolution of the properties of dusty star-forming galaxies in the redshift range $z\simeq [0-2.5]$  in the absence of systematic observational biases. Section 6 summarizes our conclusions. 
Throughout this paper we assume a $\Lambda$CDM cosmology with $H_0=70 \: \rm Km\:s^{-1} 
 Mpc^{-1}$, $\Omega_0=0.27$,  $\Omega_\Lambda=0.73$ and $\sigma_8=0.8$. 

\section{The data}

\begin{figure}
\begin{center}
\includegraphics[scale=0.4]{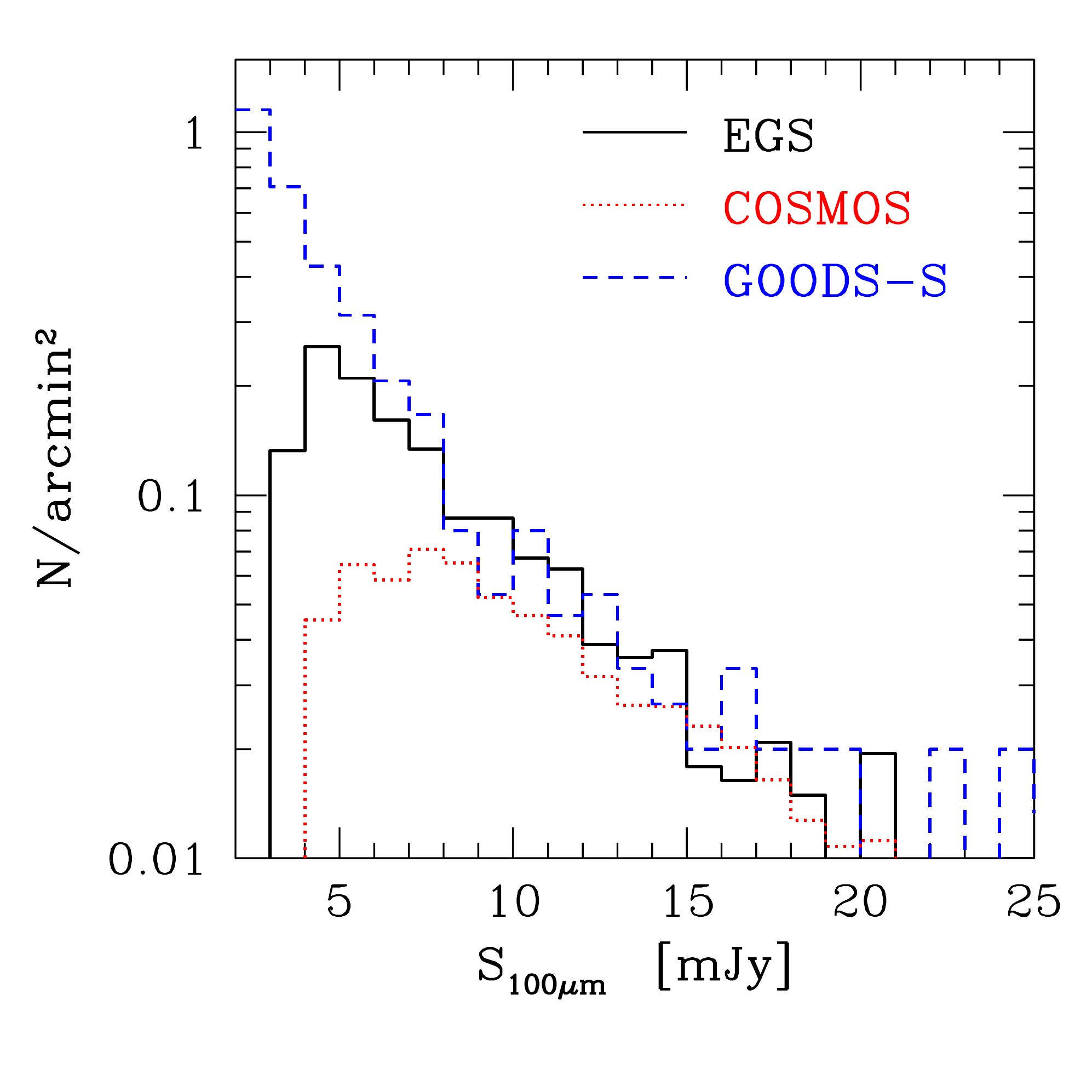}
\caption{Normalized distribution of PEP sources as a function of 100$\mu$m flux in the Extended Groth Strip (EGS), COSMOS and GOODS-South fields.
\label{fig:fluxes}}
\end{center}
\end{figure}

The COSMOS and Extended Groth Srip (EGS) regions have been observed by the PACS (Poglitsch et al. 2010) instrument onboard the {\it Herschel} Space Observatory (Pilbratt et al. 2010) as a part of the PACS Evolutionary Probe (PEP,  D. Lutz et al. 2011) Survey, aimed at  studying the properties and cosmological evolution of the infrared population up to redshifts $z\sim 3-4$. 
 We refer to the Lutz et al. (2011) and Berta et al. (2010)  papers for further information on the survey and fields.

The total number of sources detected at 100$\mu$m in the blind (i.e. without the use of priors, in order to minimize the possible bias effects introduced by a double selection) catalogues at the  $\simgt 3\sigma$ confidence level is  5360 in the COSMOS region and 1012 in the EGS.  The corresponding 100$\mu$m flux limits are $\sim 3$ mJy for COSMOS and $\sim 2$ mJy for EGS. We note that in this work we will only make use of observations taken at 100$\mu$m since, as it was shown by Magliocchetti et al. (2011), clustering results obtained at 160$\mu$m were virtually identical to those derived at 100$\mu$m. 

The distribution of sources as a function of flux in these two fields is shown in Figure \ref{fig:fluxes}, and it is compared with that of 100$\mu$m-selected galaxies observed by PEP in the GOODS-South (GOODS-S), to date the region covered by the deepest  {\it Herschel}  data. As it is possible to appreciate from the Figure, the EGS field is  $\simgt 80$\% complete for fluxes  $S_{100 \mu\rm{m}}\ge 5$ mJy, while COSMOS only reaches the above completeness level for $S_{100 \mu\rm{m}}\ge 8$ mJy. These figures are also in agreement with those estimated from simulations (cfr. Lutz et al. 2011 and Berta et al. 2010). 

\begin{figure*}
\begin{center}
\includegraphics[scale=0.4]{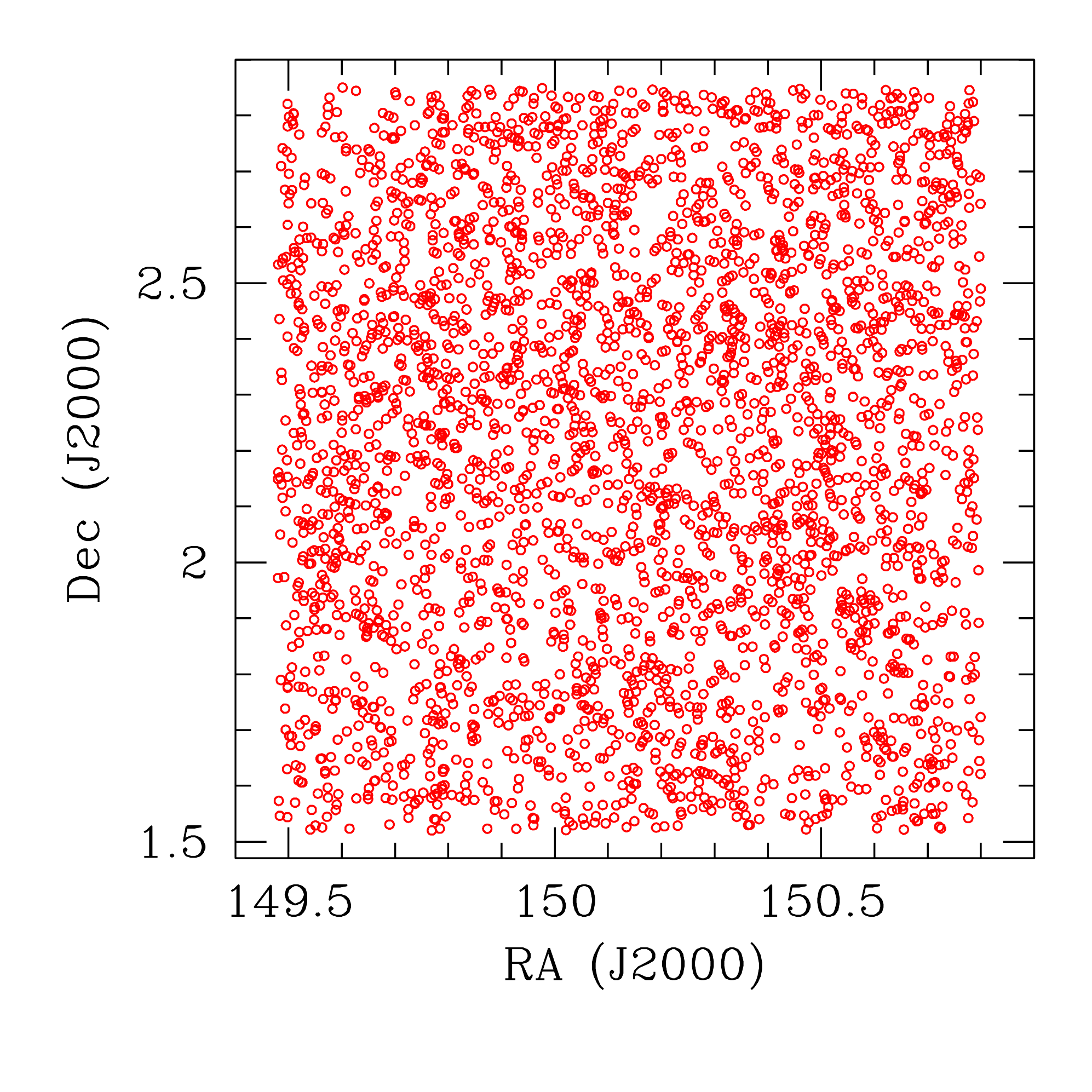}
\includegraphics[scale=0.4]{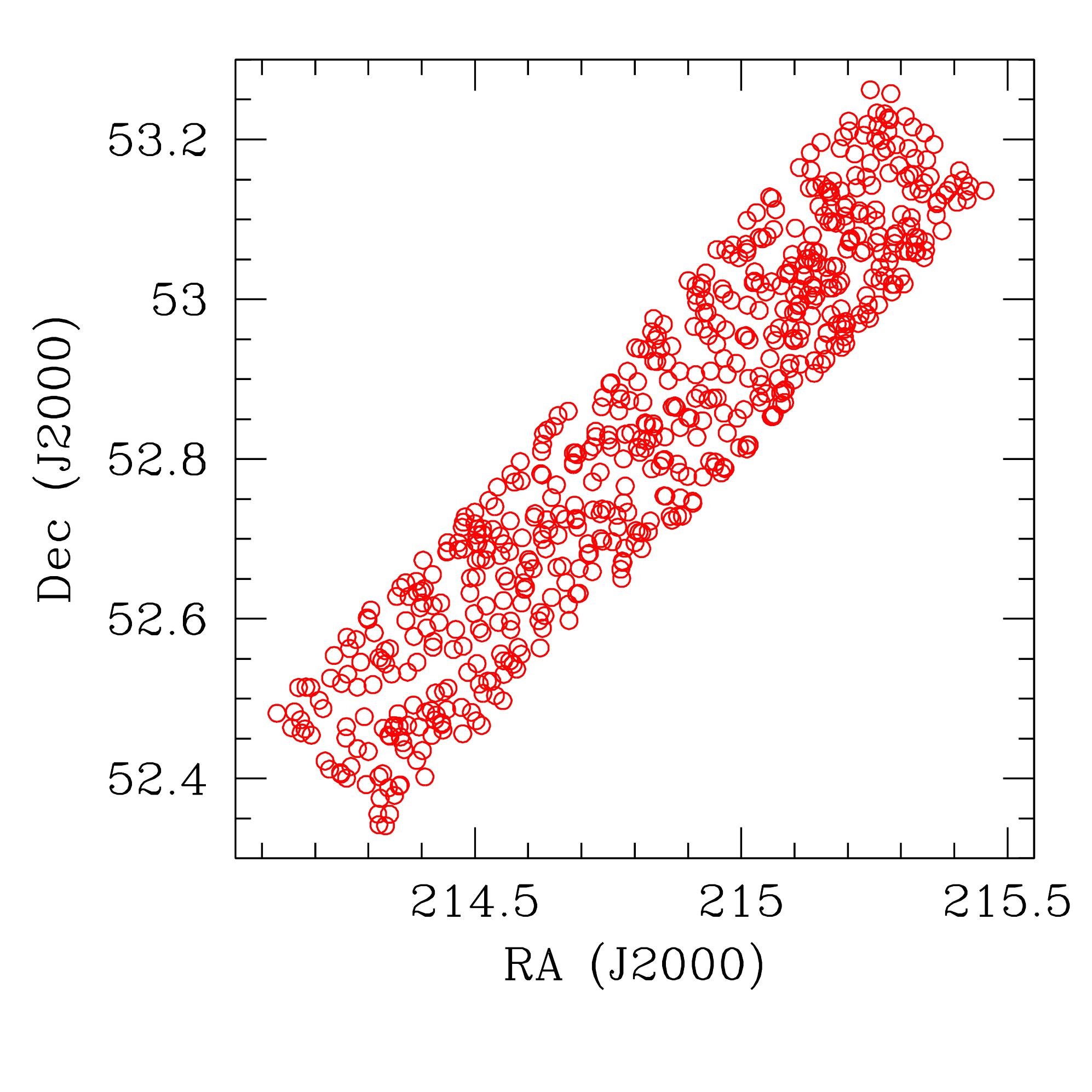}
\caption{Projected distribution of sources for the fields considered in our analysis. Left-hand side is for COSMOS, while the right-hand side represents the EGS. In order to ensure $\simgt 80$\% completeness, the first field only includes sources brighter than $S_{100 \mu\rm{m}}=8$ mJy, while the second one is obtained for a flux-cut $S_{100 \mu\rm{m}}=5$ mJy. 
\label{fig:fields}}
\end{center}
\end{figure*}

Since clustering studies need high completeness levels, in the following analysis we will only concentrate 
on those sources belonging to the blind PEP catalogues which have 100$\mu$m PACS fluxes above the 
$\sim 80$\% completeness level. Applying the above flux cuts, the number of sources are respectively 
3913 in COSMOS  ($S_{100 \mu\rm{m}}\ge 8$ mJy) and 751 in EGS  ($S_{100 \mu\rm{m}}\ge 5$ mJy). Their projected distributions onto the sky are shown in the two panels of Figure \ref{fig:fields}.

\section{Clustering Analysis}

\subsection{The Angular Correlation Function}

\begin{figure*}
\begin{center}
\includegraphics[scale=0.4]{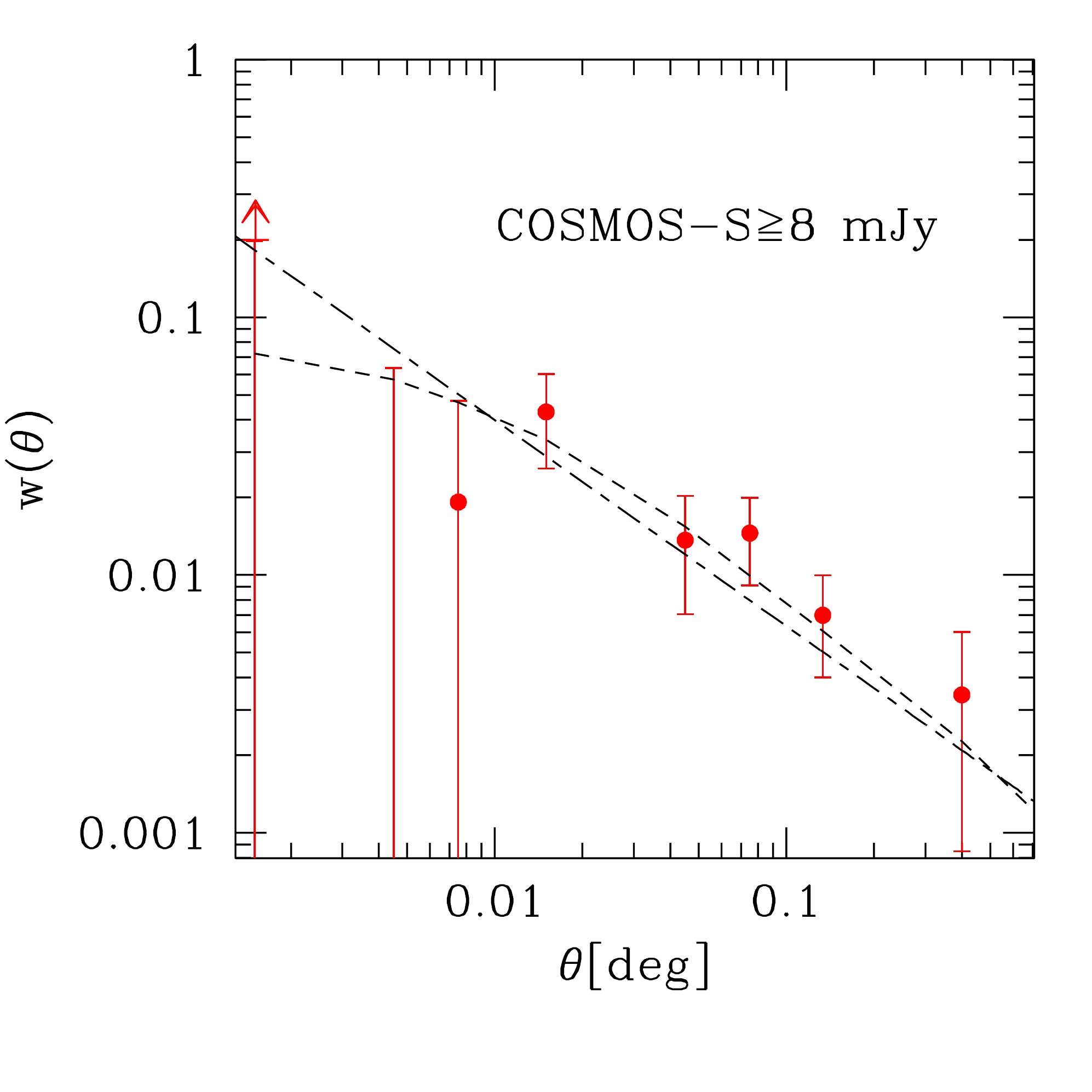}
\includegraphics[scale=0.4]{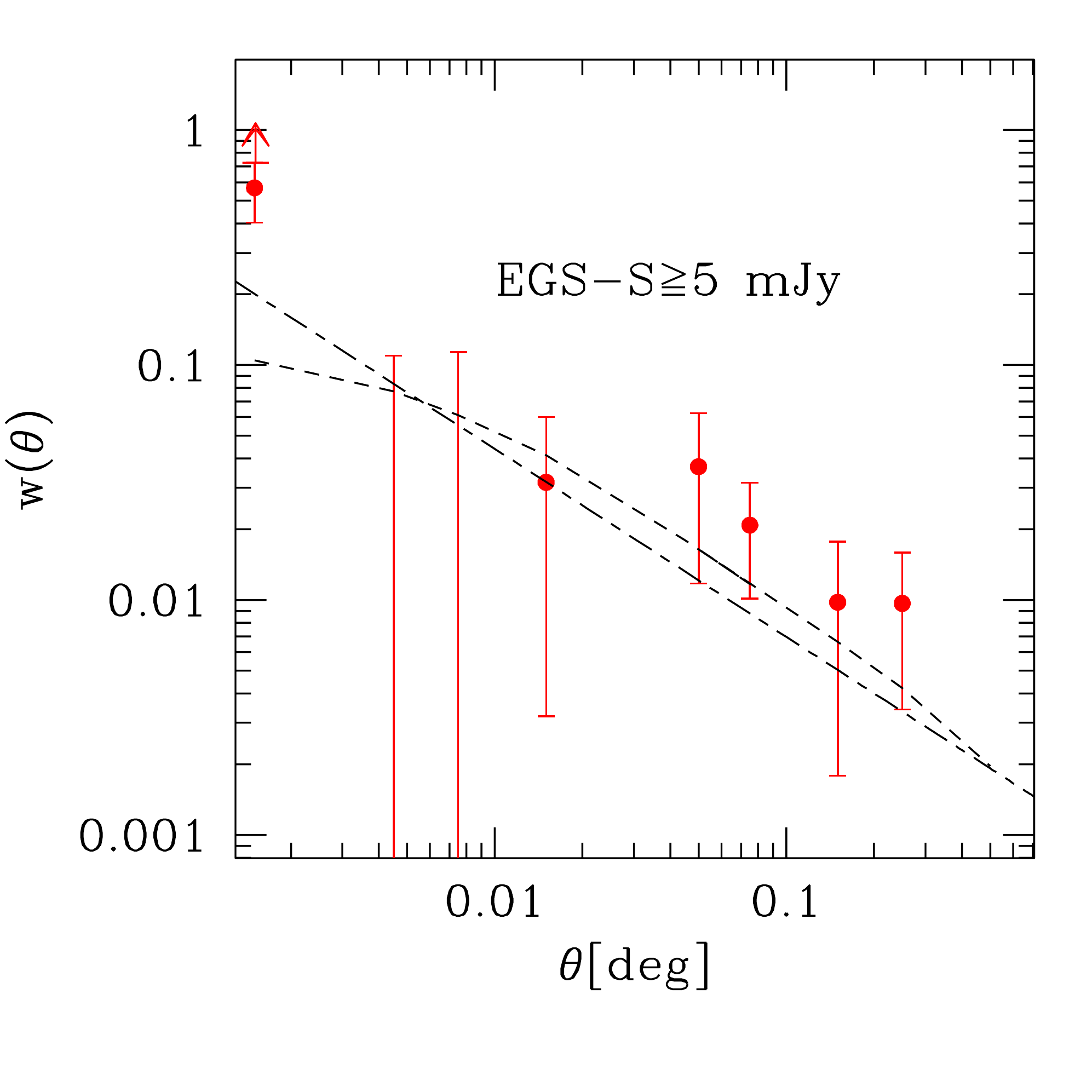}
\caption{Projected two-point correlation function for 100-$\mu$m-selected sources in COSMOS ($S_{100 \mu\rm{m}}\ge 8$ mJy, left-hand panel) and EGS ($S_{100 \mu\rm{m}}\ge 5$ mJy, right-hand panel). 
The short-long dashed lines show the best power-law fit, obtained for $\gamma=1.8$ (fixed) and for different values of the amplitude $A$, while the dashed lines are the best fits to models obtained for a varying minimum halo mass (see text and Table 1 for details). 
\label{fig:w_theta}}
\end{center}
\end{figure*}

The angular two-point correlation function
$w(\theta)$ gives the excess probability, with respect to a
random Poisson distribution, of finding two sources in the solid
angles $\delta\Omega_1$ $\delta\Omega_2$ separated by an angle
$\theta$. In practice, $w(\theta)$ is obtained by comparing the
observed source distribution with a catalogue of randomly
distributed objects subject to the same mask constraints as the
real data. \\
We chose to use the estimator (Hamilton 1993)
\begin{eqnarray}
w(\theta) = 4\times \frac{DD\cdot RR}{(DR)^2} -1, 
\label{eq:xiest}
\end{eqnarray}
where $DD$, $RR$ and $DR$ are the number of data-data, random-random 
and data-random pairs separated by a distance $\theta$. 

Given the requirement for $\ge80$\% completeness in all the considered samples (cfr Section 2) and the considerable uniformity of data sampling throughout the fields, 
we have estimated $w(\theta)$ by simply generating random catalogues of about 100K sources so to cover the whole surveyed areas minus the outermost regions which showed irregular data coverage. 
Within the considered areas, the number of 100$\mu$m-selected sources is 3005 in the COSMOS field and 647 in the EGS. 

 $w(\theta)$ in eq. (\ref{eq:xiest}) for the above objects was then estimated on angular scales ranging from 
$10^{-3}$ degrees up to values of $\theta\sim 0.1-0.5$ degrees, depending on the geometry of the two fields
since the upper limit cannot be larger than about half of the maximum scale probed by a survey.
 We also note that we did not correct for incompleteness as a function of IR flux when counting pairs, as its effect would be much smaller than the large uncertainties associated with the estimates of $w(\theta)$, due to the relatively small number of sources belonging to the considered fields.\\

Figure \ref{fig:w_theta} then shows our results for the angular correlation function in  COSMOS ($S_{100 \mu\rm{m}}\ge 8$ mJy) on the left-hand panel and EGS ($S_{100 \mu\rm{m}}\ge 5$ mJy) on the right-hand panel.  Due to the angular resolution of the Herschel mirror at the considered 
wavelength ($\sim 8^{\prime\prime}$) which would cause source blending on lower angular scales, 
the results in the smallest $\theta$ bins of the angular correlation function  merely represent lower limits and therefore have been indicated with upward arrows.

 The error-bars have been obtained  from jack-knife resampling. In the first case, the COSMOS field was divided into 9 quadrants of approximately the same area and the correlation function $w(\theta)$ was calculated for nine realizations of the data, obtained each time by omitting one quadrant. Errors were then obtained from the variance in $w$. The same approach was used on the EGS, where the whole field was divided into 10 quadrants.

If we then assume the standard power-law form for $w(\theta)=A\theta^{1-\gamma}$,
we can estimate the parameters $A$ and $\gamma$ by using a least-squares
fit to the data.  Given the large errors in $w$,  
in all cases we chose to fix $\gamma$ to the standard value $\gamma=1.8$.
Although somewhat arbitrary, this figure and its assumed lack of dependence 
on redshift is justified by Large Scale Structure observations of large enough samples of 
high redshift sources so to allow for a direct estimate of the slope of $w(\theta)$
at different look back times (e.g. Porciani, Magliocchetti \& Norberg 2004; 
Le Fevre et al. 2005).  The small areas of both the COSMOS and EGS fields introduce a negative bias through the integral constraint $\int w^{\rm est} d \Omega_1 d\Omega_2=0$. We allow for this by fitting to $A \theta^{1-\gamma}-C$, where $C= 1.93$ in the case of COSMOS and $C=2.54$ for EGS were found by numerical integration as in Roche \& Eales (1999).

The long-short dashed lines in Figure  \ref{fig:w_theta}  represent the best power-law fits to all the various data-sets. The associated best-fit values for the amplitude are: 
$A$($S_{100 \mu\rm{m}}\ge 8$ mJy)=$[1.0\pm 0.3]\cdot 10^{-3}$ for the COSMOS sample and
 $A$($S_{100 \mu\rm{m}}\ge 5$ mJy)=$[1.1^{+0.7}_{-0.6}]\cdot 10^{-3}$ for the EGS sample. 
They are summarized in Table 1.
The above values for the clustering amplitude are comparable with those obtained by Magliocchetti et al. (2011) for the sample of $S_{100 \mu\rm{m}}\ge 2$ mJy-PEP selected galaxies in the GOODS-S field. We also note that the low clustering signal observed by Maddox et al. (2010) for galaxies selected at 250$\mu$m  within the Herschel-ATLAS consortium is compatible with our PEP results.

\subsection{Relation to spatial quantities}

\begin{figure}
\begin{center}
\includegraphics[scale=0.4]{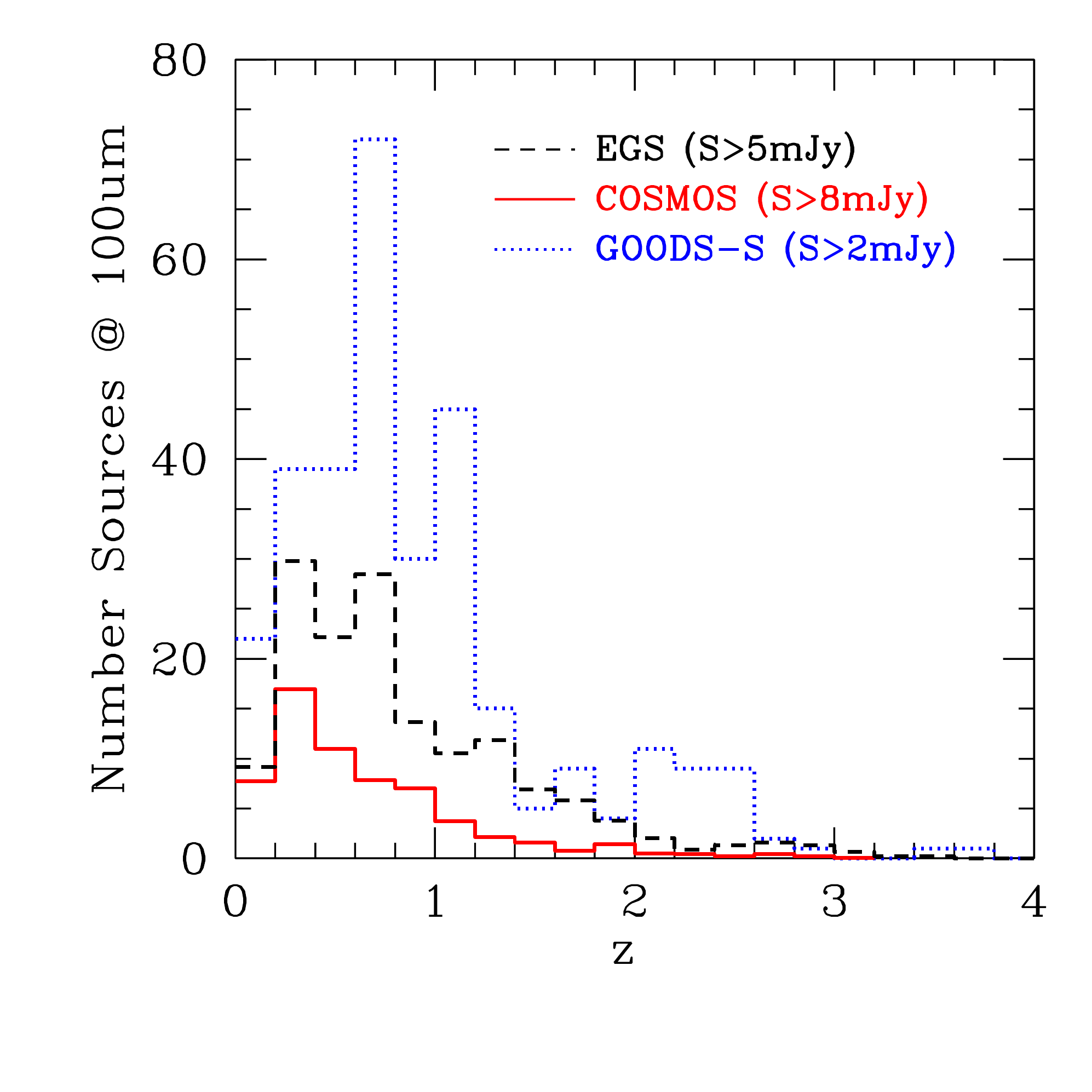}
\caption{ Normalized redshift distribution of 100$\mu$m-selected sources in the GOODS-South field 
($S_{100 \mu\rm{m}}\ge2$ mJy),
for the EGS field  ( $S_{100 \mu\rm{m}}\ge5$ mJy) and COSMOS  ($S_{100 \mu\rm{m}}\ge8$ mJy).
\label{fig:N_z}}
\end{center}
\end{figure}

The standard way of relating the angular two-point correlation
function $w(\theta)$ to the spatial two-point correlation function
$\xi(r,z)$ is by means of the relativistic Limber equation (Peebles,
1980):
\begin{eqnarray}
w(\theta)=2\:\frac{\int_0^{\infty}\int_0^{\infty}F^{-2}(x)x^4\Phi^2(x)
\xi(r,z)dx\:du}{\left[\int_0^{\infty}F^{-1}(x)x^2\Phi(x)dx\right]^2},
\label{eqn:limber} 
\end {eqnarray}
where $x$ is the comoving coordinate, $F(x)$ gives the correction for
curvature, and the selection function $\Phi(x)$ satisfies the relation
\begin{eqnarray}
{\cal N}=\int_0^{\infty}\Phi(x) F^{-1}(x)x^2 dx=\frac{1}{\Omega_s}
\int_0^{\infty
}N(z)dz,
\label{eqn:Ndense} 
\end{eqnarray}
in which $\cal N$ is the mean surface density on a surface of solid angle
$\Omega_s$ and $N(z)$ is the number of objects in the given survey
within the shell ($z,z+dz$). 

Both COSMOS and EGS have reliable and statistically complete catalogues of source redshifts (either spectroscopic or photometric; D. Rosario and O.Ilbert private communication). 
The redshift distributions of PEP-selected sources brighter respectively than 5 mJy and 8 mJy in these two fields are shown in Figure~\ref{fig:N_z}. They are compared with the $N(z)$ derived for GOODS-S sources brighter than  $S_{100 \mu\rm{m}}=2$ mJy presented in Magliocchetti et al. (2011). \\
Two main features can be observed in this plot: the first one is a systematic trend for brighter sources to be found at increasingly lower redshifts
($<z>\simeq 0.65$ for sources brighter than 8 mJy, $<z>\simeq 0.85$ for $S_{100 \mu\rm{m}}\ge 5$ mJy and 
$<z>\simeq 0.92$ for $S_{100 \mu\rm{m}}\ge 2$ mJy).  The second noticeable feature is a progressive disappearance of the $1.7\simlt z\simlt 2.6$ population with brighter limiting fluxes. We will discuss this second point and its consequences on the clustering results in more detail in Section 5.

If we then adopt a spatial correlation function of the form $\xi(r,z)=(r/r_0)^{-1.8}$ and we consider the 
redshift distributions at the various flux limits as presented in Figure~\ref{fig:N_z}, from the observed angular correlation function and for the adopted cosmology we obtain:  $r_0^{100\mu \rm m}$ (S$\ge$8 mJy)=$4.3^{+0.7}_{-0.7}$~Mpc and
 $r_0^{100\mu \rm m}$(S$\ge$ 5 mJy)=$5.8^{+1.8}_{-2.0}$~Mpc,
where both quantities are comoving and the first result refers to COSMOS sources, while the second one to the EGS sample.  We remark that the choice for a fixed value of the quantity $\gamma=1.8$, although a sensible one for describing the data presented in Figure 3, implies that in the process of data-fitting the covariance between the amplitude of the correlation function and its slope has been ignored. This implies that the errors on $r_0$ presented above might have been underestimated.

From the above results it appears a tendency for brighter sources to be less clustered than fainter ones. This trend  is confirmed if one goes to even fainter fluxes and includes the results of Magliocchetti et al. (2011) for $S_{100 \mu\rm{m}}\ge2$ sources in the GOODS-S field ($r_0^{100\mu \rm m}=6.3^{+1.1}_{-1.3}$~Mpc).

We also note that the results for the brightest (and on average closest, cfr Figure \ref{fig:N_z}) infrared galaxies are virtually indistinguishable from those obtained locally for star-forming objects selected via various methods. Indeed, the COSMOS correlation length mirrors those derived for optically-selected late-type galaxies (e.g.  $r_0=5.2\pm 0.4$ Mpc; Madgwick et al. 2003 - $r_0=4.5\pm 0.2$; Zehavi et al. 2011), for UV-selected galaxies both locally ($r_0=4.6\pm 0.6$ Mpc at $z\sim 0.2$; Heinis et al. 2007) and at higher redshifts ($r_0=4.9\pm 0.5$ Mpc at $z\sim 0.9$; Heinis et al. 2007) and for IRAS-selected sources ($r_0=5.4\pm 0.2$ Mpc; Saunders, Rowan-Robinson \& Laurence 1992).

Our local results are also in good agreement with the very recent estimates of van Kampen et al. (2012),
 obtained for a local ($z\le 0.3$) sample of 250$\mu$m-selected galaxies taken from the {\it Herschel}-ATLAS survey so to posses reliable spectroscopic redshifts ($r_0=5.62\pm 1.14$ Mpc), although their result is 
likely boosted by the presence of a structure found at $z= 0.164$. Indeed the value reported by van Kampen et al. (2012) for the correlation length in the lowest redshift bin reads $r_0=3.29\pm1.10$ Mpc.

\begin{figure}
\begin{center}
\includegraphics[scale=0.4]{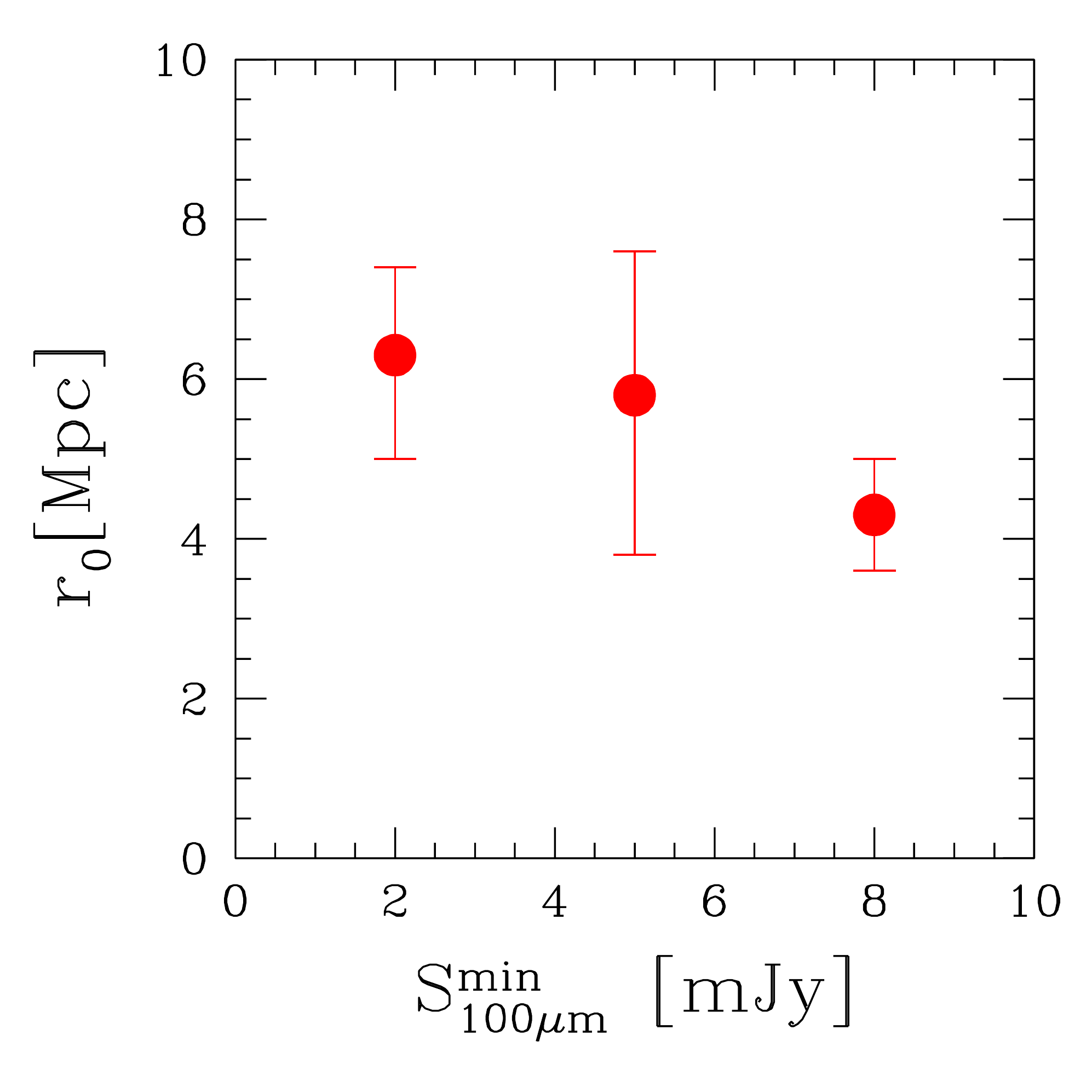}
\caption{Comoving correlation length as a function of 100$\mu$m flux as seen in various fields observed by the {\it Herschel}-PEP survey. The point at 8 mJy is obtained for COSMOS sources, while that at 5 mJy is derived for objects belonging to the EGS field. The result at 2 mJy is taken from Magliocchetti et al. (2011) for sources in the GOODS-S field.
\label{fig:r0vsflux}}
\end{center}
\end{figure}

A visual representation of the dependence of the clustering length $r_0$ on 100$\mu$m flux is provided in Figure \ref{fig:r0vsflux}. A more detailed discussion of its implications will be presented in Section 5.

\section{Connection with physical properties: the Halo Bias Approach}

\begin{table*}
\begin{center}
\caption{ Properties of the fields observed by the PEP survey at 100 $\mu$m and included in our analysis: a) field name;  b) limiting flux for $\ge 80$\% completeness;  c) total source number above 80\% completeness; d) amplitude $A$ of the angular two-point correlation function $w(\theta)$; e) comoving correlation length $r_0$; f) minimum dark halo mass $M_{\rm min}$.}
\begin{tabular}{cccccc} 
  Field Name    & S$^{\rm min}_{100\mu \rm m}$ [mJy]& Number sources & $A$ & $r_0$ [Mpc] & Log$_{10}$ M$_{\rm min} $ [M$_\odot$] \\
\hline
\hline
COSMOS& 8& 3005& $(1.0^{+0.3}_{-0.3} ) \times 10^{-3}$&  $4.3^{+0.7 }_{-0.7} $&$11.6^{+0.4 }_{-0.7}$\\
EGS& 5& 647& $(1.1^{+0.7 }_{-0.6} ) \times 10^{-3}$&  $5.8^{+1.8 }_{-2.0} $&$12.4^{+0.4 }_{-0.6}$\\
\hline
\hline

\end{tabular}
\end{center}
\end{table*}

\begin{figure}
\begin{center}
\includegraphics[scale=0.4]{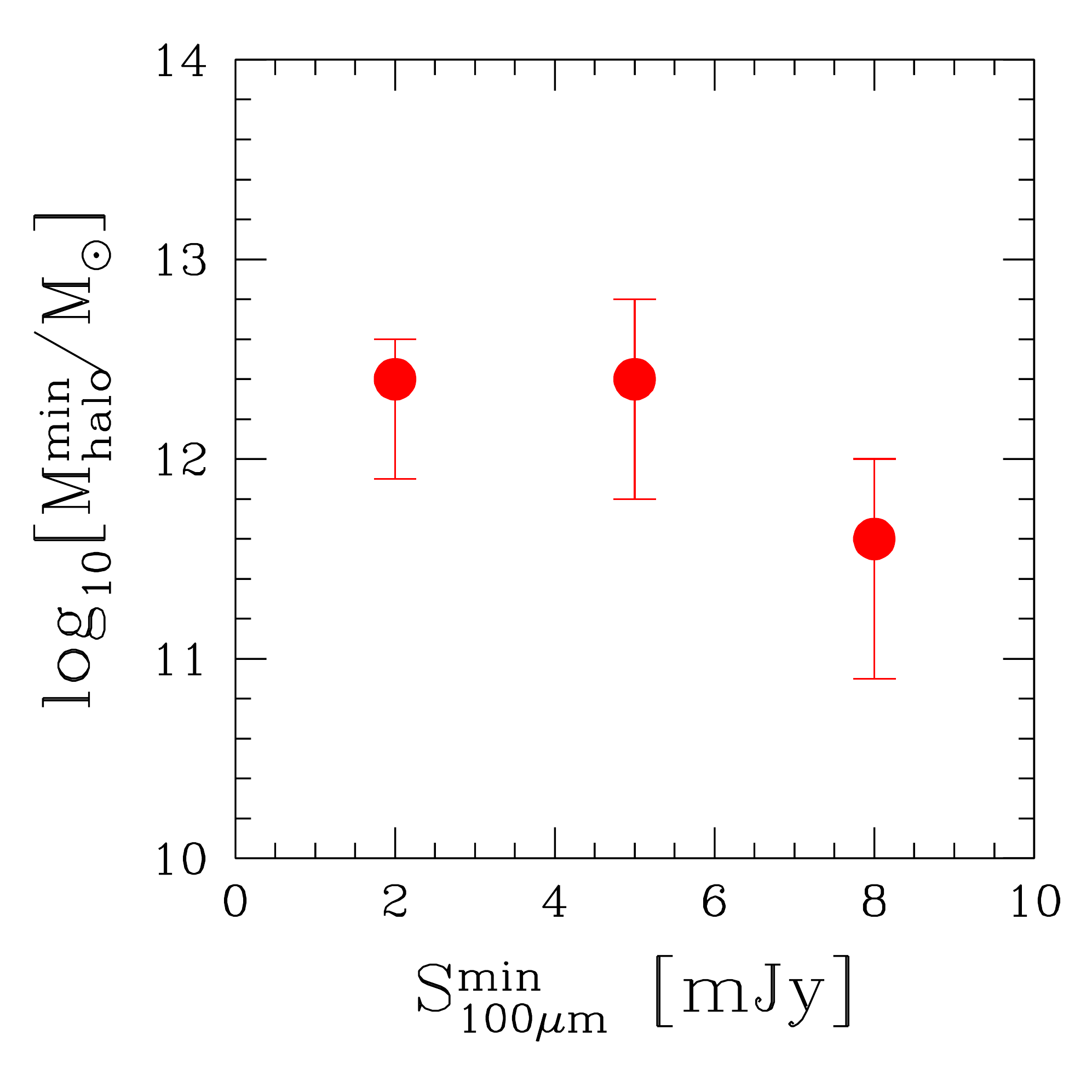}
\caption{Minumum dark halo mass as a function of 100$\mu$m flux as seen in various fields observed by the {\it Herschel}-PEP survey. The point at 8 mJy is obtained for COSMOS sources, while that at 5 mJy is derived for objects belonging to the EGS field. The result at 2 mJy is taken from Magliocchetti et al. (2011) for sources in the GOODS-S field.
\label{fig:massvsflux}}
\end{center}
\end{figure}

The easiest way to relate the clustering trend of a population of extra-galactic sources with their physical properties is by means of the Halo Bias approach (Mo \& White 1996; Sheth \& Tormen 1999) 
which describes the observed clustering signal in terms of the product between the two-point correlation function of the underlying dark matter distribution $\xi_ {\it dm}(r,z)$ and the square of the so-called 
bias function $b(M_{\rm min},z)$,  which at a given redshift solely depends on the minimum mass $M_{\rm min}$ of the haloes in which the detected astrophysical objects reside via the relation: 
\begin{equation}
\xi(r,z)=\xi_ {\it dm}(r,z) \cdot b(M_{\rm min},z)^2.
\label{eq:bias}
\end{equation}

More sophisticated models have been introduced in the recent years in order to provide a more realistic description of the observed clustering properties of extra-galactic sources.
One of the most successful ones, the so-called Halo Occupation Model (HOM, e.g. Scoccimarro et al. 2001) relates the clustering properties of a chosen population of galaxies to the way such objects populate their dark matter haloes, i.e. relaxes the somehow unrealistic assumption, implicit in the halo bias model,  of having a one-to-one correspondence between a galaxy and its dark matter halo. \\
 However desirable,  such an analysis is unfortunately not possible on our datasets due to a relatively poor signal-to-noise ratio. In fact, the HOM is a three parameter model: together with the minimum mass of the dark matter halo it also depends on the the quantities $\alpha$ and $N_0$ according to the expression 
\begin{eqnarray}
N(M) = N_0(M/M_{\rm min})^{\alpha} \;\;\;
\rm{if}\ M \ge  M_{\rm min}, \nonumber
\label{eq:Ngal}
\end{eqnarray}
where $N(M)$ is the number of galaxies within a halo of some mass $M$.  And a three-parameter fit will only be possible when larger fields will be included in the analysis.
For the time being, we notice that the quantities $M_{\rm min}$ and $\alpha$ are covariant, so that within the HOM framework, higher values for $\alpha$ in general correspond to lower 
values for $M_{\rm min}$. 
This implies that in the presence of multiple halo occupancy, the values for $M_{\rm min}$ 
found within the HOM scenario will be (slightly) lower than those obtained via the halo bias model adopted in this work. 

Having made this necessary digression, we can then go back to the Halo Model framework and 
compute the theoretical angular two-point correlation function $w(\theta)_{\rm th}$ predicted by this model 
starting from eq.(\ref{eq:bias}) and projecting it once again by following eq.(\ref{eqn:limber}) with the $N(z)$'s 
 provided in Figure \ref{fig:N_z}.\\
The best values for $M_{\rm min}$ at the two limiting fluxes of 5 mJy and 8 mJy were then found by a $\chi^2$-fit to the observed clustering signal. This gives:\\
 ${\rm Log_{10}}[M_{\rm min}/M_\odot]$($S\ge 8$ mJy)=$11.6^{+0.4}_{-0.7}$ and
${\rm Log_{10}}[M_{\rm min}/M_\odot]$($S\ge 5$ mJy)=$12.4^{+0.4}_{-0.6}$, 
where again the first result refers to the COSMOS field, while the second one to the EGS sample.  These results are summarized in Table 1. The corresponding best-fit $w(\theta)_{\rm th}$ are shown in Figure \ref{fig:w_theta}  by the dashed curves. 

The values obtained in our analysis, especially those derived for the brightest and, as already stressed, on average closest sources, are again in excellent agreement with results found in the literature for the typical masses of halos locally hosting star-forming galaxies. Just to mention a  few, Heines et al. (2007) quote bias levels which, by using the Sheth \& Tormen (1999) bias model,  convert to minimum halo masses of  $
{\rm Log_{10}}[M_{\rm min}/M_\odot]\sim 10.7$ and ${\rm Log_{10}}[M_{\rm min}/M_\odot]\sim 11.5$,
 respectively for their low-redshift and $z\simeq 0.9$ samples. And while the above figures were derived for UV-selected galaxies, very similar results are also obtained by Magliocchetti \& Porciani (2003) when analysing the clustering properties of a sample of star-forming galaxies extracted from the 2dF survey (${\rm Log_{10}}[M_{\rm min}/M_\odot]\sim 11$) and also by Zehavi et al. (2011) for their sample of blue galaxies drawn from the seventh release of the SDSS  (${\rm Log_{10}}[M_{\rm min}/M_\odot]\sim 11$).

Furthermore, although we have not made use of the HOM model to estimate our best-fit parameters, we can still rely on its theory to draw some conclusions on the sources which produce the clustering signal presented in Figure \ref{fig:w_theta}.  In fact, both the estimated angular correlation functions, irrespective of the particular field considered in this work or of the flux cut,  present a flattening (coinciding in most cases with negative values of $w(\theta)$) on angular scales $0.004 \simlt  \theta \simlt 0.01$ degrees. 
This can be easily explained as a further evidence for the well known "avoidance behaviour" of local star-forming galaxies which prefer to stand isolated and not have companions within the same dark matter halo (cfr Magliocchetti \& Porciani 2003 or Zehavi et al.  2011 just to mention a couple of works).

As a last point, we show in Figure \ref{fig:massvsflux} the dependence of the minimum halo mass $M_{\rm min}$ on 100$\mu$m flux. As in the case of the estimated correlation lengths, there is a trend for  sources brighter than 8 mJy to be hosted in smaller halos.  This is also further confirmed by the Magliocchetti et al. (2011) results which can be used to extend the analysis to even fainter, $S_{100 \mu\rm{m}}\ge2$, sources.   
The next Section will discuss the implications of both this and other findings of our work.

\section{The Redshift Evolution of Clustering of FIR Galaxies}

\subsection{Dealing with redshift dependence}

 In the previous sections we have found a tendency for the clustering signal of PEP sources selected at 100$\mu$m to decrease as one moves to higher flux limits. This in turn implies that brighter galaxies are hosted by smaller haloes, with values ranging from $\sim 10^{11.5} M_\odot$ at the brightest fluxes probed by our analysis up to $\sim 10^{12.5} M_\odot$  for a $\le 5$ mJy flux cut. In Section 3.2 we have also learned that, in the case of 100$\mu$m-selected galaxies, higher fluxes  on average imply more local sources (cfr. Figure \ref{fig:N_z}). Putting together these two pieces of information in a straightforward way then would imply a clustering strength which increases for increasing redshifts, at least up to $z\sim 1$. But is that so?

As already extensively discussed, Figure \ref{fig:N_z} clearly shows that the contribution of the high, $z\sim 2$, population to the total number of PEP sources, basically negligible for $S_{100\mu\rm m}\simgt 8$ mJy,  tends to steeply increase when moving to fainter fluxes. 
Magliocchetti et al. (2011) have shown this $z\sim 2$ population to be a very extreme one, being very strongly clustered ($r_0\sim 19$ Mpc) and hosted in very massive/cluster-like halos ($M\simgt 
10^{13.5} M_{\odot}$).  So the question one might want to ask is: how much of the measured clustering signal for fluxes $\simlt 8$ mJy is solely due to the high-redshift population and how much comes from more local sources? In other words: can we separate within our data-sets the clustering signal stemming from high-z sources from that originating from low-redshift ones?

Indeed, this is possible. In fact, in our case, the observed correlation function produced by  the two low-z ($A$) and high-z ($B$) populations can be written as:
\begin{eqnarray}
\xi(r,z)=
\left \{ 
\begin{array} {ccc}
\xi_A(r,z) &  {\rm for}& z<1.7\\
 \xi_B(r,z) &  {\rm for} &1.7\simlt z\simlt 2.6, 
\end{array}
\right .
\label{eq:xinew}
\end{eqnarray}
so that the observed projected correlation function $w(\theta)$ will be
\begin{eqnarray}
w(\theta)=n_A^2w_A(\theta)+n_B^2w_B(\theta)
\label{eq:wnew}
\end{eqnarray}
(see e.g. Magliocchetti et al. 1999), where $n_A$ and $n_B$ are the relative space densities of the two populations at the different flux cuts, normalized so that $n_A+n_B=1$. Note that, since we want to investigate the clustering behaviour of objects located at different redshifts, the cross-correlation signal $n_An_B \left[w_A(\theta)w_B(\theta)\right]^{1/2 }$ in equation (\ref{eq:wnew}) is zero.

\begin{figure}
\begin{center}
\includegraphics[scale=0.4]{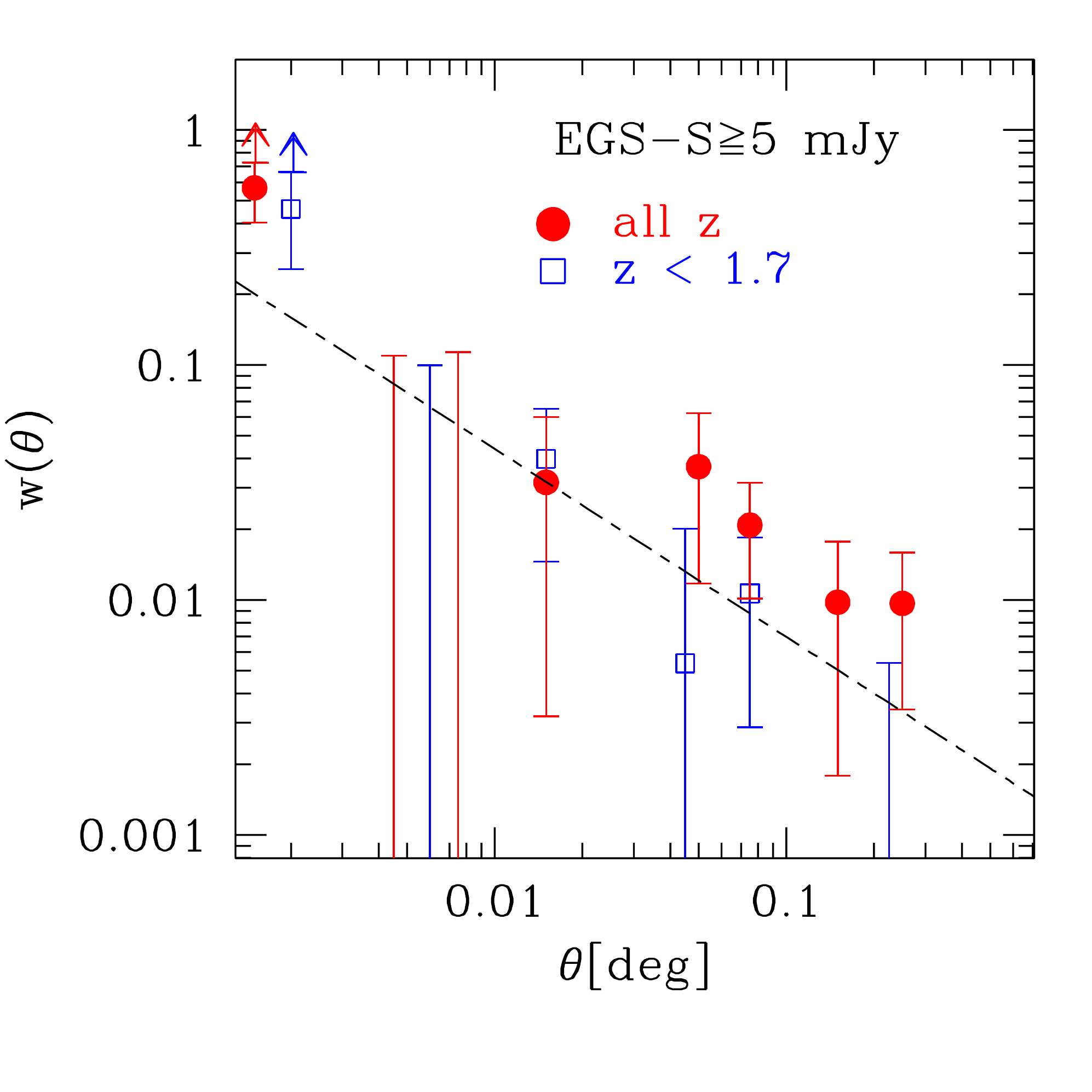}
\caption{ Projected two-point correlation function for 100-$\mu$m-selected sources in the EGS field  with $S_{100 \mu\rm{m}}\ge 5$ mJy and $z<1.7$ (blue, open squares) compared to that produced by all sources in the EGS subject to the same flux cut (red, filled dots). The short-long dashed line shows the best power-law fit  obtained for the more local sample by assuming $\gamma=1.8$ (see text and Table 2 for details) 
\label{fig:wegslowz}}
\end{center}
\end{figure}

\begin{table*}
\begin{center}
\caption{Properties of a homogeneous sample of star-forming galaxies observed by the {\it Herschel}-PEP and IRAS surveys at $\lambda \sim 60\mu$m - rest frame. The first column reports the field name and the considered redshift range. S$_{\rm min}$ is  the mJy flux cut,  $r_0$ the correlation length and  M$_{\rm min} $ the logarithm of the minimum halo mass.  $<z>$ indicates the mean redshift of each sample,  $<\lambda_{\rm RF}>$ the rest-frame wavelength, while  $<$L$_{\rm min}>$ and $<$SFR$_{\rm min}>$ respectively indicate the average minimum bolometric luminosity and the average minimum star formation rate.  The first five lines are the results of the analysis presented in this work, while the sixth one refers to the sample of $160\mu$m-selected galaxies in the GOODS-S field as obtained by Magliocchetti et al. (2011) and the last one to the results obtained by Saunders et al. (1992) for IRAS galaxies. The redshift distribution for IRAS galaxies was taken from Saunders et al. 1990.   The superscript "$1$" illustrates the clustering results obtained by removing the high-$z$ clustering signal by means of equations \ref{eq:xinew} and \ref{eq:wnew}, while  "$2$" those derived from a direct analysis of the $z< 1.7$ sample (see text for details).   }
\begin{tabular}{llllllll} 
{\small  Field Name }   & S$_{\rm min}$& $r_0$ [Mpc] & M$_{\rm min} $ [M$_\odot$]& $<z>$&$<\lambda_{\rm RF}>$[$\mu$m]& $<$L$_{\rm min}>$[L$_\odot$] &$<$SFR$_{\rm min}>$ [M$_\odot$/yr]\\
\hline
\hline
COSMOS[$100\mu\rm m$ ]  ($z<1.7$)& 8&  $4.1^{+0.8 }_{-1.0} $& $11.1^{+0.4 }_{-0.7}$&$0.56\pm 0.36$&$67 \pm 14$& $11.6\pm0.6$&$72\pm 106$\\
EGS[$100\mu\rm m$ ] ($z<1.7$)$^1$& 5&  $5.0^{+2.2 }_{-3.3} $&$11.9^{+0.5 }_{-1.1}$&$0.68\pm 0.39$&$62 \pm 14$& $11.6\pm 0.5$&$72\pm 84$ \\
EGS[$100\mu\rm m$ ] ($z<1.7$)$^2$& 5&  $5.1^{+1.4 }_{-1.8} $&$11.8^{+0.6 }_{-1.9}$&$0.68\pm 0.39$&$62 \pm 14$& $11.6\pm 0.5$&$72\pm 84$ \\
GOODS-S[$100\mu\rm m$ ]  ($z<1.7$)$^1$& 2&  $5.1^{+1.8 }_{-2.6} $&$12.0^{+0.4 }_{-0.7}$&$0.70\pm 0.35$&$61\pm 13$& $11.2\pm 0.5$&$29\pm 27$ \\
GOODS-S [$100\mu\rm m$] ($z<1.7$)$^2$& 2&  $4.3^{+1.8 }_{-2.3} $&$11.5^{+0.9 }_{-3.0}$ &$0.70\pm 0.35$&$61\pm 13$& $11.2\pm 0.5$&$29\pm 27$ \\
GOODS-S[$160\mu\rm m$] ($1.7\le z\le 2.6$)& 5&  $17.4^{+2.8 }_{-3.1} $&$13.7^{+0.3 }_{-0.4}$ &$2.1\pm 0.3$&$52\pm 5$& $12.1\pm 0.2$&$226\pm 113$ \\
IRAS[$60\mu\rm m$] ($z\sim0$)&600& $5.4^{+0.2 }_{-0.2} $&$11.4^{+0.2}_{-0.2}$&$\sim$0.02 &60&$11.0 \pm 0.4$ &$18\pm 17$\\
\hline
\hline

\end{tabular}
\end{center}
\end{table*}

We take the correlation function $\xi_B$ of 100$\mu$m-selected high-z galaxies  from Magliocchetti et al. (2011) and estimate the values of $n_A$ and $n_B$ from the redshift distributions presented in Section 3.2. Specifically: 
$\xi_B=(r/r_0)^{-1.8}$, with $r_0=19^{+2.6 }_{-2.9}$; $n_A(S\ge 8\;\rm mJy)\simeq 0.96$, $n_B(S\ge 8\;\rm mJy)\simeq 0.04$; $n_A(S\ge 5\;\rm mJy)\simeq 0.91$, $n_B(S\ge 5\;\rm mJy)\simeq 0.09$; $n_A(S\ge 2\;\rm mJy)\simeq 0.9$, $n_B(S\ge 2\;\rm mJy)\simeq 0.1$. 

Once again assuming a power-law form with $\gamma=1.8$ for the spatial correlation function of low-redshift sources, the above values lead to clustering lengths respectively of:\\ 
$r_0(z< 1.7)=4.1^{+0.8 }_{-1.0} $, for $S\ge 8$ mJy (only slightly changed from the previous analysis); \\
$r_0(z < 1.7)=5.0^{+2.2 }_{-3.3} $,  for $S\ge 5$ mJy;\\
$r_0(z <1.7)=5.1^{+1.8 }_{-2.6} $,  for $S\ge 2$ mJy;\\
and minimum halo masses of:\\
$\rm Log_{10}$[M$_{\rm min}$/M$_\odot$]= $11.1^{+0.4 }_{-0.7}$, for $S\ge 8$ mJy; \\
$\rm Log_{10}$[M$_{\rm min}$/M$_\odot$]=$11.9^{+0.5 }_{-1.1}$, for $S\ge 5$ mJy;\\
$\rm Log_{10}$[M$_{\rm min}$/M$_\odot$]=$12.0^{+0.4 }_{-0.7}$, for $S\ge 2$ mJy.\\
All the above values are summarized in Table 2.

In all cases, although somewhat masked by the large uncertainties, the net result is a decrement of both the correlation length and of the halo mass of the host. Such a decrement is increasingly more pronounced as the fraction of high-z FIR galaxies increases.

In order to cross-check the reliability of the above results,  in the EGS field we have also directly examined   
the clustering properties of the 592 $S_{100\mu \rm m} \ge 5$ mJy galaxies endowed with either a spectroscopic or a photometric redshift determination, $z < 1.7$. This is presented in Figure \ref{fig:wegslowz} by the (blue) open squares and compared with the filled (red) points which reproduce the $w(\theta)$ obtained for EGS galaxies at all redshifts. The amplitude decrement, already appreciable by eye, implies in this case a correlation length 
$r_0=5.1^{+1.4 }_{-1.8} $ and a minimum mass for the host halo $\rm Log_{10}$[M$_{\rm min}$/M$_\odot$]=$11.8^{+0.6}_{-1.9}$, in excellent agreement with the previous determinations. \\
A similar approach performed on $z<1.7$ GOODS-S galaxies with $S_{100\mu \rm m}\ge 2$ mJy leads to 
$r_0=4.3^{+1.8 }_{-2.3}$ and M$_{\rm min}= 11.5^{+0.9 }_{-3.0}$ [M$_\odot$]. 
Despite the uncertainties, the values just found are once again in excellent agreement with those derived by making use of equations \ref{eq:xinew} and \ref{eq:wnew}.

\subsection{Results and comparisons with other works}

The values obtained in Section 5.1 and Table 2 indicate that there is a substantial constancy in the clustering strength of relatively local, $z < 1.7$, sources as a function of their 100$\mu$m flux limit. 

\begin{figure*}
\begin{center}
\includegraphics[scale=0.4]{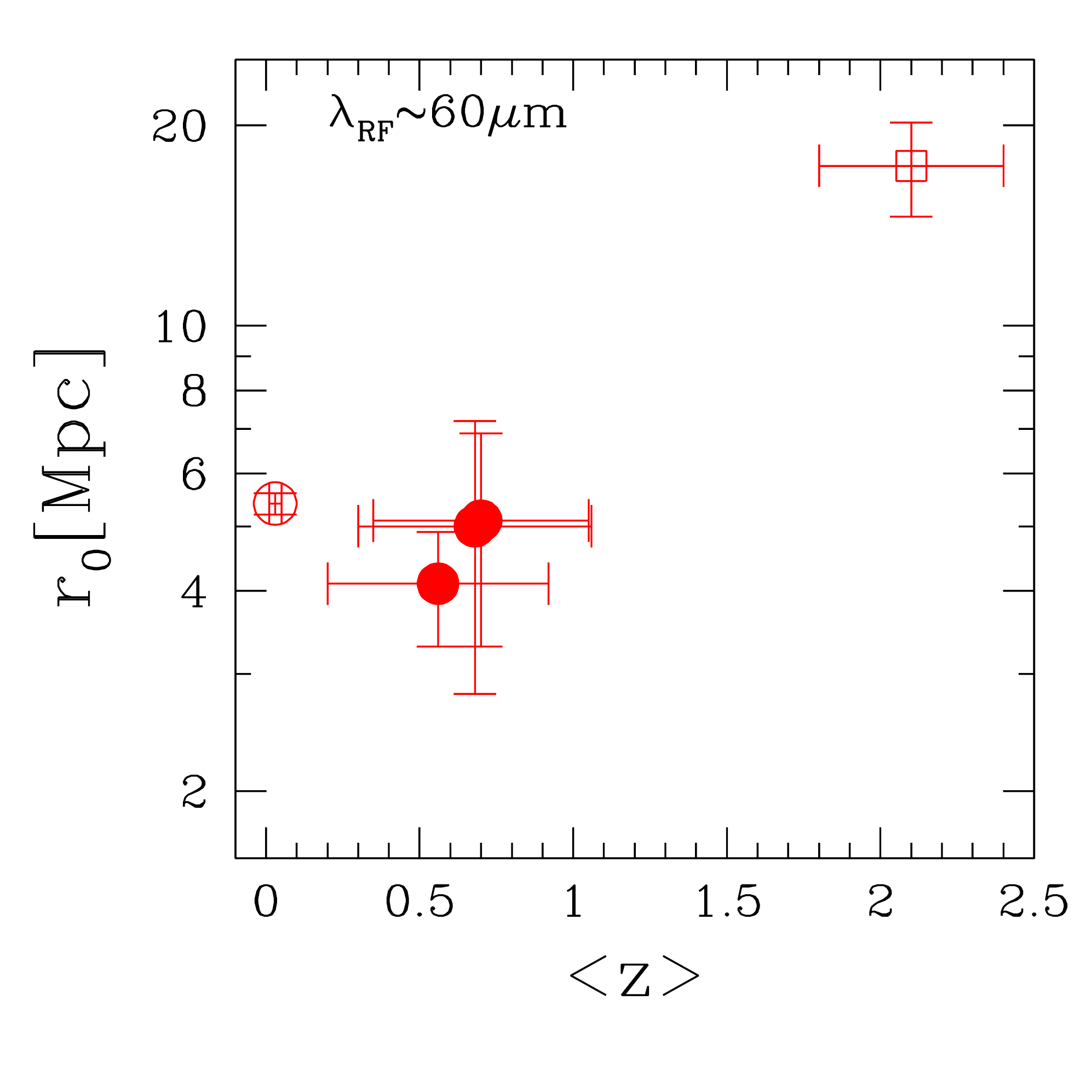}
\includegraphics[scale=0.4]{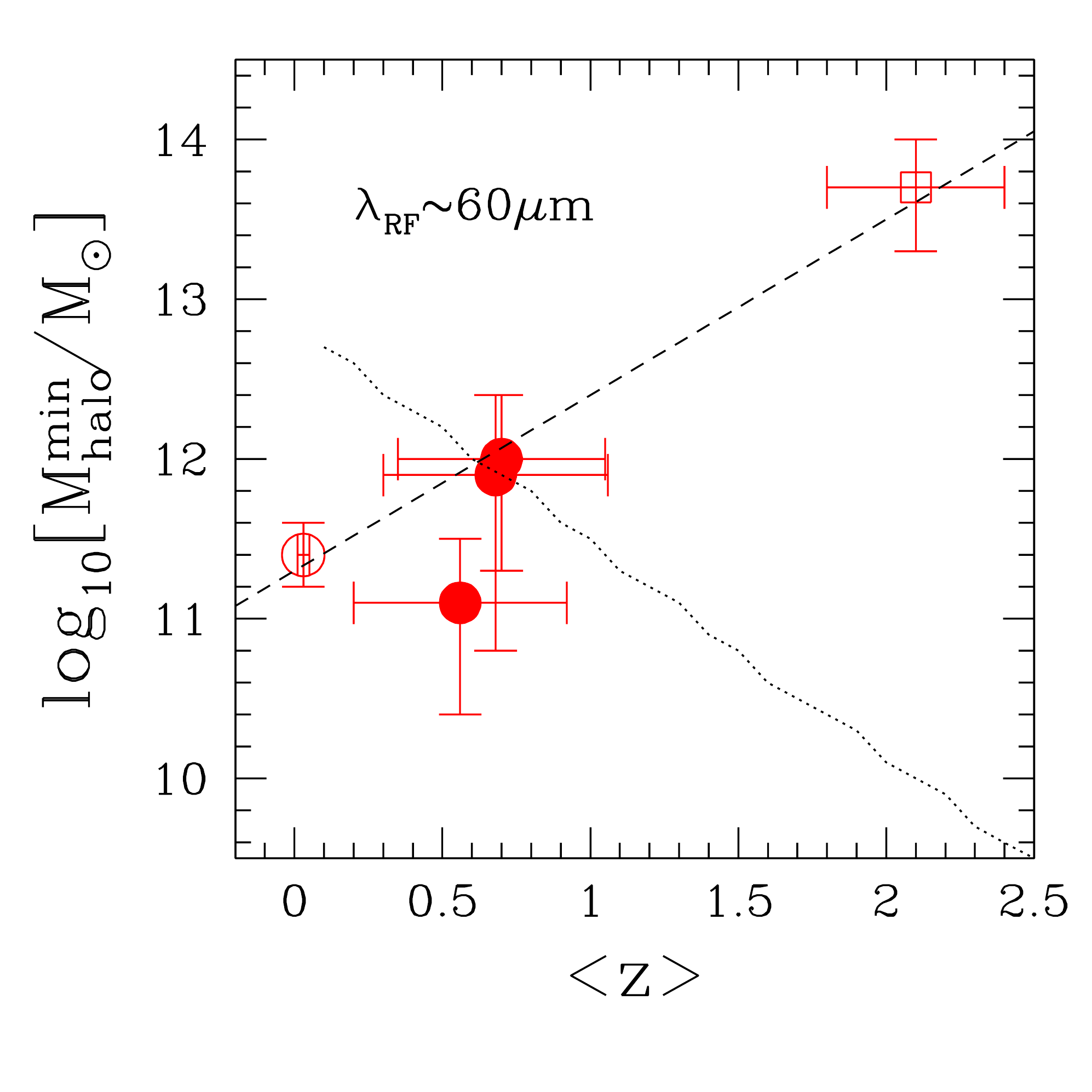}
\caption{Evolution of the clustering length $r_0$ (left-hand panel) and of the minimum halo mass M$_{\rm min}$(right-hand panel) for a sample of galaxies selected at about the same, $\lambda\sim 60\mu$m, rest-frame wavelength. 
The filled points show the results obtained in this work, while the empty squares are taken from Magliocchetti et al. (2011) and are derived for a sample of $z\simeq 2$ GOODS-S galaxies selected at 160$\mu$m. The open dots are the Saunders et al. (1992) result for IRAS galaxies. In the right-hand panel, the dashed line represents the best fit to the data obtained for a functional form Log[M$_{\rm min} /\rm M_\odot]=\alpha \cdot z+\beta$, with $\alpha= 1.1$ and $\beta=11.3$, while the dotted line represents the redshift evolution of M$_{\sigma}$ (see text for details).
\label{fig:res3}}
\end{center}
\end{figure*}

We can actually extend the above analysis to create a homogeneous sample of galaxies selected at the same rest-frame wavelength in order to study the clustering evolution of star-forming galaxies free from the usual biases connected with the selection band. In fact, as it is possible to appreciate from the first five lines of Table 2, 
the 100$\mu$m-selected samples at the various flux cuts and for $z < 1.7$ identify a redshift range (where the limits correspond to the rms of the distributions presented in Figure \ref{fig:N_z}), which in turn corresponds to a range of rest-frame wavelengths all centred at $\lambda\sim 60\; \mu$m. It is then possible to compare such results with those obtained in the nearby universe for IRAS galaxies. This is done in Figure \ref{fig:res3}, which reports the redshift evolution of the clustering strength of 60$\mu$m-selected galaxies both by looking at their correlation length (left-hand panel) and minimum halo mass (right-hand panel). The filled points indicate the results of our work, while the open circle in the left-hand panel is the result for $r_0$ obtained by Saunders et al. (1992), once the slope $\gamma$ was changed from their measured value to the $\gamma=1.8$ used throughout this work. The corresponding point in the right-hand panel of Figure \ref{fig:res3} for the minimum halo mass of IRAS galaxies was estimated with our code, to reproduce the correlation length reported by Saunders et al. (1992). The high, $z\sim 2$, redshift points, illustrated in Figure \ref{fig:res3} by the open squares, represent the values for the correlation length and  
M$_{\rm min}$ obtained by Magliocchetti et al. (2011) for a sample of {\it Herschel}-PEP galaxies, selected in the redshift range $z=[1.7-2.6]$ at $160\mu$m in the GOODS-S field.  In fact, in the considered redshift range, an observed $\lambda=160\mu$m corresponds to a rest-frame value of $ 52\pm 5 \mu$m (cfr Table 2), perfectly compatible with our chosen sample. 

What we have at hand is therefore a sample which includes star-forming galaxies all selected at the same rest-frame frequency. 
Furthermore, most of the sources belonging to this sample have comparable bolometric luminosities and therefore star formation rates (SFR henceforth). In fact, we can use the redshift distributions  from Section 3.2 to estimate the mean minimum bolometric luminosity which corresponds to the various flux cuts $S_{\rm min}$  via
\begin{eqnarray}
<L_{\rm min}>=\frac{\int^{z_{\rm max}} _{z_{\rm min}} L'_{\rm min}(z) N(z)\; dz }{\int^{z_{\rm max}}_{z_{\rm min}} N(z)\; dz}, 
\end{eqnarray}
with 
 \begin{eqnarray}
L'_{\rm min}(z)=\int l_{\rm min}(\lambda^*,z)\; f(\lambda) \; d\lambda, 
\end{eqnarray}
where $l_{\rm min}(\lambda^*,z)=4\pi S_{\rm min} d_L^2/K(\lambda^*,z)$, is the minimum monochromatic luminosity at $\lambda^*$ corresponding to $S_{\rm min}$, $d_L$ is the luminosity distance and  the K-correction is  expressed  as $K(\lambda^*,z)=(1+z)f(\lambda^*/(1+z))/f(\lambda^*)$. 
$f$ is the normalized emission spectrum of the source.  This was chosen in agreement with the results obtained by Gruppioni et al. (2010) for  the population mixture of PEP-selected galaxies, to be of the kind generated by a moderate star-forming source (M82-like spectrum)  at low-to-intermediate redshifts and as produced by an intense star-forming activity (Arp220-like spectrum) at $z\sim 2$. However, we stress that the following results do not greatly vary if we assume the same spectrum for all objects.

The results of the above  analysis are summarized in Table 2, which clearly shows that the values for the averaged minimum bolometric luminosities obtained at relatively low redshifts at the various flux cuts coincide within the errors, and, in addition, that the luminosity of 160$\mu$m-selected sources in GOODS-S at $z\sim 2$ is comparable with those derived for 100$\mu$m-selected,  $z< 1.7$ sources in the COSMOS and EGS fields. By converting such luminosities into star formation rates with the usual formula 
$\rm SFR [\rm M_\odot/yr]=1.8\times 10^{-10} L_{\rm bol}/L_\odot$ (Kennicutt et al. 1998), we then obtain that 
the high-z sample includes sources which are as active as  low-redshift COSMOS and EGS galaxies, i.e. {\it  the sources included in our analysis are not only selected at $\sim$ the same rest-frame wavelength, but also show comparable luminosities and star-formation rates}. In other words, the sample of PEP-selected galaxies presented in this part of the work is reasonably {\it free from both selection and luminosity biases, so that the evolution of their clustering properties can be considered as only driven by the intrinsic evolution of the galaxy population. }/

We note that the IRAS sample has been included in our analysis even though its sources are in general fainter than those observed by {\it Herschel} (even though with luminosities which are comparable to those of the $z < 1.7 $ GOODS-S sample).  This is because it has been shown (e.g. Hawkins et al. 2001) that their clustering properties do not depend on their luminosity (while they do depend on their colours), even in the most extreme cases.  The Hawkins et al. (2001) conclusions are also confirmed by our analysis which shows, within the errors, a remarkable consistency of the clustering properties of $z\simlt 1$ dusty star-forming galaxies with SFR. 

Figure 8 then shows the redshift evolution of the clustering properties of sources with comparable absolute luminosities and star formation rates and which are selected at  roughly the same wavelength at the various cosmological distances.   There is a clear trend for the clustering strength of these galaxies to increase at earlier epochs. This implies that also the masses of the hosts tend to increase as one moves to higher redshifts, showing a very clear evidence for downsizing. 
However, while the rise of both the correlation length and the halo mass is relatively limited up to $z\sim 1$, so much that it could be considered as almost constant within the uncertainties, there is a jump of about a factor 3 in $r_0$ and of a couple of orders of magnitude in the halo mass when moving from $z\simeq 1$ to $z\simeq 2$. 
The best linear fit to the data presented in the left-hand panel of Figure 8 is provided by a functional form
$\rm Log_{10}$[M$_{\rm min}/{\rm M}_\odot]= \alpha z+\beta$, with $\alpha=1.1\pm 0.2$ and $\beta=11.3\pm 0.2$ and it is illustrated by the dashed line.

Also, while the halo masses of FIR-selected galaxies in the relatively local universe all lie below M$_\sigma$\footnote {M$_\sigma$ is the typical mass-scale at which a 1$\sigma$ fluctuation is collapsing.}  (represented in the right-hand panel of Figure 8 by the dotted line),  meaning they are  relatively common objects, the halo mass found at $z\sim 2$ (M$_{\rm min}$ as large as $10^4$M$_\sigma$) corresponds to extremely high density (and therefore rare) peaks of the underlying density field.  Furthermore, Magliocchetti et al. (2011), by estimating the number density of dark matter halos via straightforward integration of the Sheth \& Tormen (1999) mass function found that "every dark matter halo which reaches the mass to host a (dusty) star-forming galaxy at redshift $z\sim 2$ will be inhabited by (at least) one of such objects". 

This, and the results obtained so far in this work, imply that intense star-forming activity  within galaxies takes place in very different environments at the different cosmological epochs. At relatively low, $z< 1.7$ redshifts, the hosts of such systems are low-mass and isolated galaxies, and only a fraction (which decreases as one approaches the more local universe)  of the virialized halos of comparable mass will be actively forming stars. On the other hand, the very same amount of star formation found at higher redshifts happens in extremely massive/proto-cluster like structures, and is presumably triggered by interactions of close pairs. Furthermore, this process is very common, as approximately all the $z\sim 2$ halos which reach the threshold mass to host a star-forming galaxy will be inhabited by at least one of such objects.

\section {conclusions}
We have used data taken by the {\it Herschel}-PEP survey at 100$\mu$m in the COSMOS and Extended Groth Strip fields to investigate the possible dependence of the clustering properties of FIR sources on their fluxes. To this aim, we have selected two samples of galaxies respectively brighter than 8 mJy (COSMOS) and 5 mJy (EGS) and estimated their projected angular two-point correlation function $w(\theta)$. 

By fitting the available observations to the functional form $w(\theta)=A\theta^{1-\gamma}$ (with $\gamma$ fixed to the value $\gamma=1.8$) and by then de-projecting the best-fit values by means of the self-consistent redshift distributions in the two fields, we find for the correlation length: $r_0(S_{100 \mu\rm{m}}\ge 8$ mJy) =$4.3^{+0.7}_{-0.7}$ Mpc for COSMOS sources and  $r_0(S_{100 \mu\rm{m}}\ge 5$ mJy) =$5.8^{+1.8}_{-2.0}$ Mpc for EGS galaxies.   These data show a mild trend for the clustering strength of FIR-selected sources to {\it decrease} with {\it increasing} fluxes. Such a trend is confirmed if we add to the above results those obtained by Magliocchetti et al. (2011) for PEP galaxies brighter than 2 mJy in the GOODS-South field ($r_0(S_{100 \mu\rm{m}}\ge 2$ mJy) =$6.3^{+1.1}_{-1.3}$ Mpc). 

A similar (inverse) dependence on the flux limit is found for the minimum mass of the dark-matter halos hosting such FIR-selected sources. By applying the extended Press \& Schechter theory (Sheth \& Tormen 1999), we derive from our data:  ${\rm Log_{10}}[M_{\rm min}/M_\odot]$($S_{100 \mu\rm{m}}\ge 8$ mJy)=$11.6^{+0.4}_{-0.7}$ and ${\rm Log_{10}}[M_{\rm min}/M_\odot]$($S_{100 \mu\rm{m}}\ge 5$ mJy)=$12.4^{+0.4}_{-0.6}$.\\

We also report a remarkable flattening of the observed $w(\theta)$ for angular distances $0.004\simlt \theta\simlt 0.01$ degrees. Such a trend, which proceeds down to the smallest probed scales in the case of sources  
with fluxes $S_{100 \mu\rm{m}}\ge 8$ mJy, is due to the well known "avoidance behaviour" of local star-forming galaxies which prefer to stand isolated and not to have companions within the same dark matter halos (cfr Magliocchetti \& Porciani (2003) or Zehavi et al.  (2011)). \\

\noindent
But what is the physical explanation for the above findings? 
A careful investigation of the redshift distributions of PEP sources selected at 100$\mu$m shows that the effect of lowering the flux limit is basically two-fold: on the one hand it moves the peak of the distribution to (slightly) higher redshifts, while more importantly it affects the appearance and relative importance of a population of $z\sim 2$ sources. Such a population makes about 10\% of the total counts for $S_{100 \mu\rm{m}}\ge 2$ mJy, while it already becomes negligible for sources brighter than 8 mJy ($\sim 4$\%).

Magliocchetti et al. (2011) have shown this $z\sim 2$ population to be very strongly clustered, with correlation lengths as high as $r_0\sim 19$ Mpc and corresponding halo masses M$_{\rm min}\sim 10^{13.9}$ M$_\odot$, so that it is plausible to assume these objects responsible for the enhancement of the clustering strength of faint,  $S_{100 \mu\rm{m}}< 8$ mJy FIR sources. We have then sub-divided our samples into intermediate ($z <1.7$) redshift galaxies and higher-z ($z\simeq [1.7-2.6]$) ones and applied two different methods to evaluate the clustering properties of intermediate redshift FIR galaxies: one which relied on the subtraction of the clustering signal produced by the high redshift population from the total measured angular correlation function, and one which directly estimated $w(\theta)$ for  $z<1.7$, 100$\mu$m -selected galaxies in the GOODS-S and EGS fields.  Both methods converge, indicating an overall constancy of the clustering strength of relatively local sources, independent of flux cuts. \\

\noindent
With the above results in hand, we have then studied the {\it intrinsic} redshift evolution of the clustering properties of dusty star-forming galaxies. This was done by extending our analysis to comprise sources in the redshift range [0-2.5] which: a) were all selected at the same rest-frame wavelength to minimize selection biases; b) showed comparable luminosities and therefore star-formation rates in order to overcome Malmquist-like bias effects. \\
To achieve this, we added to our set of $z<1.7$ 100$\mu$m-selected sources both data from the clustering analysis of $z\sim 0$, 60$\mu$m-selected galaxies from the IRAS survey (Saunders et al. 1992) and also the results from Magliocchetti et al. (2011) for $z\sim 2$, 160$\mu$m selected galaxies in the GOODS-S.  All these objects were observed at approximately the same rest-frame $\lambda=60 \mu$m. Furthermore, 
the star formation rates estimated for  $z\sim 2$ sources (SFR$\simgt 100$ M$_\odot$/yr) are comparable with those  derived for $<z>\sim 0.5$ galaxies in the COSMOS and EGS fields. 

Our analysis for this homogeneously-selected sample shows a strong evolution of the clustering properties  with look-back time. More specifically, both the clustering length and the minimum halo mass can be considered as approximately constant (or at most very weakly dependent on redshift), irrespective of the chosen flux cut, up to about $z\sim 1$. Above this redshift value, there is a huge jump of a factor $\sim 3$ in $r_0$ and of about two orders of magnitude in M$_{\rm min}$. 
In this latter case, the dependence of  M$_{\rm min}$ on $z$ can be parametrized as $\rm Log_{10}$[M$_{\rm min}/{\rm M}_\odot]= \alpha z+\beta$, with $\alpha=1.1^{+0.2}_{-0.2}$ and $\beta=11.3^{+0.2}_{-0.2}$.

This and the other results obtained in this work imply that  intense star-formation within galaxies takes place in very different environments at the different epochs. In the relatively local universe,  the hosts of such systems are low-mass and isolated galaxies, with only a fraction of  them actively forming stars. On the other hand, the very same amount of star formation found at higher redshifts happens in extremely massive/proto-cluster-like structures, and is presumably triggered by interactions of close pairs (cfr Kartaltepe et al. 2012). Furthermore, this process is very common, as all the $z\sim 2$ sources which reach threshold halo masses to host a star-forming galaxy will be inhabited by at least one of such objects.

We note that such extreme  findings for the clustering properties of star-forming galaxies at redshifts $z\sim 2-3$ are not just the result of a particularly biased field or as a particular feature of the FIR realm. Indeed clustering length values which are comparable with those presented in this work and in Magliocchetti et al. (2011) are found in the same redshift range for high-SFR objects in various deep fields and at all probed wavelengths: from the UV (e.g. Magliocchetti \& Maddox 1999), to the sub-mm regime (e.g. Hickox et al. 2012), passing through the {\it Spitzer}-24$\mu$m  (e.g. Magliocchetti et al. 2008; Brodwin et al. 2008; Starikova et al. 2012) and the BzK selection (e.g. Lin et al. 2012). We will investigate the  theoretical implications of such a behaviour in a forthcoming paper.\\
\\
\\

\noindent
{\bf Acknowledgements}
We wish to thank the referee for his/her comments which greatly improved our work.
PACS has been developed by a consortium of institutes led by MPE
(Germany) and including UVIE
(Austria); KU Leuven, CSL, IMEC (Belgium); CEA, LAM (France); MPIA
(Germany); INAF-
IFSI/OAA/OAP/OAT, LENS, SISSA (Italy); IAC (Spain). This development
has been supported by the
funding agencies BMVIT (Austria), ESA-PRODEX (Belgium), CEA/CNES
(France), DLR (Germany),
ASI/INAF (Italy), and CICYT/MCYT (Spain).


\begin{thebibliography}{}
\bibitem[\protect\citename{Amblard}2011]{Amblard}
Amblard A. et al., 2011, Nature, 470, 510
\bibitem[\protect\citename{Ar}2002]{Ar}
Arnouts S. et al., 2002, MNRAS, 329, 355
\bibitem[\protect\citename{Berta}2010]{berta}
Berta S. et al., 2010, A\&A, 518, L30
\bibitem[\protect\citeauthoryear{Brod1}2007]{bro1}
Brodwin M., Gonzalez A. H., Moustakas L. A., Eisenhardt P. R., Stanford S. A., Stern D., Brown M. J. I., 2007, ApJ, 671, L000
\bibitem[\protect\citeauthoryear{Brod}2008]{bro}
Brodwin M. et al., 2008, ApJ, 687, 65
\bibitem[\protect\citename{Bundi}2005]{bundy}
Bundy K. et al., 2006, ApJ, 651, 120
\bibitem[\protect\citename{cas}2000]{cas}
Casertano S. et al. 2000, AJ, 120, 2747
\bibitem[\protect\citename{cimatti}2004]{cimatti}
Cimatti A., Daddi E., Renzini A., 2006, A\&A, 453, L29
\bibitem[\protect\citename{cowie}1996]{cowie}
Cowie L.L., Songaila A., Hu E., Cohen J.G., 1996, AJ, 112, 839
\bibitem[\protect\citename{cor}1998]{cress}
Cress C. M. \& Kamionkowski M., 1998, 297, 486
\bibitem[\protect\citename{coorayi}2010]{cooray}
Cooray A. et al., 2010, A\&A, 518, L22
\bibitem[\protect\citename{es}2009]{es}
Estrada J. et al., 2009, ApJ, 692, 265
\bibitem[\protect\citename{far2}2006]{far2} 
Farrah D. et al., 2006, ApJ,643, L139
\bibitem[\protect\citename{fou}2010]{fou} 
Foucaud A., Conselice C.J., Hartley W.G., Lane K.P., Bamford S.P., Almaini O., Bundy K., 2010, MNRAS, 406, 147
\bibitem[\protect\citeauthoryear{Griffin et al.}2010]{gr}
Griffin M.J. et al., 2010, A\&A, 518, L3
\bibitem[\protect\citeauthoryear{Grup et al.}2010]{grup}
Gruppioni C. et al., 2010, A\&A, 518, L27 
\bibitem[\protect\citename{guzzo}2000]{guzzo}
Guzzo L. et al., 2000, ASPC, 200, 349
\bibitem[\protect\citename{hickox} 2012]{hickox}
Hickox R.C. et al., 2012, MNRAS, 421, 284
\bibitem[\protect\citename{Hamilton}1993]{Hamilton}
Hamilton A.J.S., 1993, ApJ, 417, 19
\bibitem[\protect\citename{Hart}2010]{Hart}
Hartley W.G. et al., 2010, MNRAS, 407, 1212
\bibitem[\protect\citename{Hawkins}2001]{Hawk}
Hawkins E., Maddox S., Branchini E., Saunders W., 2001, MNRAS, 325, 589
\bibitem[\protect\citename{Heavens}2004]{Heavens}
Heavens A., Panter B., Jimenez R., Dunlop J., 2004, Nature, 428, 625
\bibitem[\protect\citename{Heinis}2007]{Heinis}
Heinis S. et al., 2007, APJS, 173, 503
\bibitem[\protect\citeauthoryear{Kartaltepe}2012]{Kar}
Kartaltepe J.S. et al., 2012, ApJ, 757, 23
\bibitem[\protect\citename{Kennicutt}1998]{Kennicutt}
Kennicutt R.C.Jr. et al., 1998, ApJ, 498, 181
\bibitem[\protect\citename{Lin}2012]{Lin}
Lin L., et al., 2012, arXiv:1111.2135
\bibitem[\protect\citename{Le Fevre}2005]{Lef}
Le Fevre O. et al., 2005, A\&A, 439, 877
\bibitem[\protect\citename{Lutz}2011]{Lutz}
Lutz D. et al., 2011, A\&A, 532, 90
\bibitem[\protect\citename{Madau}1996]{Madau}
Madau P., Ferguson H.C:, Dickinson M.E., Giavalisco M., Steidel C.C., Fruchter A., 1996, MNRAS, 283, 1388
\bibitem[\protect\citename{Maddox}2010]{Maddox}
Maddox S.J. et al., 2010, A\&A, 518, L11.
\bibitem[\protect\citename{Mad}2003]{Mad}
Madgwick D. et al., 2003, MNRAS, 344, 847.
\bibitem[\protect\citename{Maglio2}1999]{Maglio2}
Magliocchetti M., Maddox S.J., 1999, MNRAS, 306, 988
\bibitem[\protect\citename{Maglio3}1999]{Maglio3}
Magliocchetti M., Maddox S.J., Lahav O., Wall J.V., 1999, MNRAS, 306, 943
\bibitem[\protect\citename{Maglio1}2007]{M1}
Magliocchetti M., Silva L., Lapi A., De Zotti G., Granato G.L., Fadda D., 
Danese L., 2007, MNRAS, 375, 1121
\bibitem[\protect\citename{Maglio}2011]{Maglio}
Magliocchetti M. et al. 2011, MNRAS, 416, 1105
\bibitem[\protect\citename{Maglio}2003]{Maglio1}
Magliocchetti M. \& Porciani C., 2003, MNRAS, 346, 186
\bibitem[\protect\citename{Maglio2}2008]{M2}
Magliocchetti M. et al., 2008, MNRAS, 383, 1131
\bibitem[\protect\citename{Mc}2006]{Mc}
McLure R.J. et al., 2006, MNRAS, 372, 357
\bibitem[\protect\citename{Mo}1996]{Mo}
Mo H.J., White S.D.M., 1996, MNRAS, 282, 347
\bibitem[\protect\citename{Peebles}1980]{Peebles}
Peebles P.J.E., 1980, {\it The Large-Scale Structure of the
Universe}, Princeton University Press
\bibitem[\protect\citename{Pil} 2010]{pil}
Pilbratt G. et al., 2010, A\&A, 518, L1
\bibitem[\protect\citename{Pog} 2010]{pog}
Poglitsch A. et al., 2010, A\&A, 518, L2
\bibitem[\protect\citename{Porci1} 2004]{porci1}
Porciani C., Magliocchetti M., Norberg P., 2004, MNRAS, 355, 1010
\bibitem[\protect\citename{Roche} 2009]{roche}
Roche N. \& Eales S. A., 1999, MNRAS, 307, 703
\bibitem[\protect\citename{Ross} 2009]{ross}
Ross N.P. et al. 2009, ApJ, 697, 1634
\bibitem[\protect\citeauthoryear{saunders et al.}1990]{saund1}
Saunders W., Rowan-Robinson M., Lawrence A., Efstathiou G., Kaiser N., Ellis R.S., Frenk C.S., 1990, 
MNRAS, 242, 318
\bibitem[\protect\citeauthoryear{saunders et al.}1992]{saund}
Saunders W, Rowan-Robinson M, Lawrence A., 1992, MNRAS, 258, 134
\bibitem[\protect\citeauthoryear{Saracco et al.}{2006}]{2006MNRAS.367..349S} 
Saracco P. et al., 2006, MNRAS, 367, 349
\bibitem[\protect\citename{Scocci}2001]{Scocci}
Scoccimarro R., Sheth R.K., Hui L.,Jain B., 2001, ApJ, 546, 20
\bibitem[\protect\citename{sheth }1999]{Sheth}
Sheth R.K., Tormen G., 1999, MNRAS, 308, 119
\bibitem[\protect\citename{shen }2007]{Shen}
Shen Y. et al., 2007, AJ, 133, 2222
\bibitem[\protect\citename{shen1 }2009]{Shen1}
Shen Y. et al., 2009, ApJ, 697, 1656
\bibitem[\protect\citename{sta}2012]{Sta}
Starikova S., Berta S., Franceschini A., Marchetti L., Rodighiero G., Vaccari M., Vikhlinin A., 2012, arXiv:1205.2045
\bibitem[\protect\citename{su}1992]{Su}
Sutherland W., Saunders W., 1992, MNRAS, 259, 413
\bibitem[\protect\citename{treu}2005]{treu}
Treu T., Ellis R.S., Liao T.X., van Dokkum P.G., 2005, ApJ, 622, L5
\bibitem[\protect\citename{Van}2012]{Van}
van Kampen et al., 2012, arXiv:1209.3213v1
\bibitem[\protect\citename{Willia}1996]{William}
Williams R.E. et al., 1996, AJ, 306, 1335
\bibitem[\protect\citename{Zehavi}2005]{Zehavi}
Zehavi I. et al., 2011, ApJ, 736, 59
\end{thebibliography}
\end{document}